\documentclass[nocopyrightspace,10pt]{sensys-proc}

\usepackage{graphicx}
\usepackage{balance}
\usepackage{comment}
\usepackage[bookmarks=false]{}
\usepackage{cite}
\usepackage{calc}

\usepackage{booktabs}
\usepackage{colortbl}
\usepackage{tabularx}
\usepackage{color}
\usepackage{xspace} 
\usepackage{graphicx,psfig,epsfig,subfigure}
\usepackage{url}
\usepackage{algorithmic}
\usepackage{algorithm}
\usepackage{epstopdf}
\usepackage{verbatim}
\usepackage{amssymb}
\usepackage{caption}
\usepackage{enumerate}
\usepackage{multicol}
\usepackage{multirow,hhline}

\numberofauthors{5}

\author{
\alignauthor Arijit Banerjee \\
        \affaddr{School of Computing}\\
        \affaddr{University of Utah}\\
       \email{arijit@cs.utah.edu}
\alignauthor Dustin Maas \\
    \affaddr{Xandem Technology}\\
    \email{dustin.314@gmail.com}
\alignauthor Maurizio Bocca \\
    \affaddr{Department of ECE}\\
    \affaddr{University of Utah}\\
    \email{maurizio.bocca@utah.edu}
    \and
    \alignauthor Neal Patwari \\
    \affaddr{Department of ECE}\\
    \affaddr{University of Utah}\\
    \email{npatwari@ece.utah.edu}\\
    \alignauthor Sneha Kasera \\
        \affaddr{School of Computing}\\
        \affaddr{University of Utah}\\
       \email{kasera@cs.utah.edu}
}

\title{Through Wall People Localization Exploiting Radio Windows}

\begin{document}

\maketitle

\begin{abstract}
We introduce and investigate the ability of an attacker to surreptitiously use an otherwise secure wireless network to detect moving people through walls, in an area in which people expect their location to be private. We call this attack on location privacy of people an ``exploiting radio windows'' (ERW) attack.
We design and implement the ERW attack methodology for through wall people localization that relies on reliably detecting when people cross the link lines
by using physical layer measurements between the legitimate transmitters and the attack receivers. We also develop a method to estimate the direction of movement of a person from the sequence of link lines crossed during a short time interval. Additionally, we describe how an attacker may estimate any artificial changes in transmit power (used as a countermeasure), compensate for these power changes using measurements from sufficient number of links, and still detect line crossings. We implement our methodology on WiFi and ZigBee nodes and experimentally evaluate the ERW attack by monitoring people movements through walls in two real-world settings.
We find that our methods achieve very high accuracy
in detecting line crossings and determining direction of motion.

\end{abstract}



\section{Introduction}
\global\long\global\long\global\long\def\dd#1{{\tfrac{\partial}{\partial#1}}}
 \global\long\global\long\global\long\def\dfd#1#2{\frac{\partial#1}{\partial#2}}
 \global\long\global\long\global\long\def\dtwod#1{\frac{\partial^{2}}{\partial#1^{2}}}
 \global\long\global\long\global\long\def\dtwodd#1#2{\frac{\partial^{2}}{\partial#1\partial#2}}
 \global\long\global\long\global\long\def\dndd#1#2#3{\frac{\partial^{#3}}{\partial{#1}\cdots\partial{#2}}}
 \global\long\global\long\global\long\def\dtwofdd#1#2#3{\frac{\partial^{2}#1}{\partial#2\partial#3}}
 \global\long\global\long\global\long\def\Var#1#2{\mbox{Var}_{#1}\left[ {#2} \right]}
 \global\long\global\long\global\long\def\Cov#1{\mbox{Cov}\left( {#1} \right)}
 \global\long\global\long\global\long\def\E#1#2{E_{#1}\left[ {#2} \right]}
 \global\long\global\long\global\long\def\I#1{\mathrm{I}_{#1}}
 \global\long\global\long\global\long\def\PR#1{P\left[{#1}\right]}
 \global\long\global\long\global\long\def\ie{{\it i.e.}}
 \global\long\global\long\global\long\def\eg{{\it e.g.}}
\global\long\global\long\global\long\def\argmin#1{\mathop{\mbox{argmin}}_{#1}}
 \global\long\global\long\global\long\def\argmax#1{\mathop{\mbox{argmax}}_{#1}}
 \global\long\global\long\global\long\def\argminX{\argmin{ X: \; X^{T}X=I_{D}}}
 \global\long\global\long\global\long\def\Order#1{\mathcal{O}\left( {#1} \right)}
 \global\long\global\long\global\long\def\ds#1{\mathrm{d}{#1}}
 \global\long\global\long\global\long\def\dv#1{\mathrm{d}\Vc{#1}}
 \global\long\global\long\global\long\def\ol#1{\overline{#1}}
 \global\long\global\long\global\long\def\tr{\mbox{tr}}
 \global\long\global\long\global\long\def\me{\mathrm{e}}
 \global\long\global\long\global\long\def\mR{\mathbb{R}}
 \global\long\global\long\global\long\def\vone{\Vc{1}}
 \global\long\global\long\global\long\def\vzero{\Vc{0}}
 \global\long\global\long\global\long\def\vI{\Vc{I}}
 \global\long\global\long\global\long\def\vbeta{\boldsymbol{\beta}}
 \global\long\global\long\global\long\def\mX{\mathcal{X}}
 \global\long\global\long\global\long\def\mY{\mathcal{Y}}
 \global\long\global\long\global\long\def\mM{\mathcal{M}}
 \global\long\global\long\global\long\def\diag#1{{\mbox{diag}\left\{  {#1}\right\}  }}
 \global\long\global\long\global\long\def\bb{\bar{b}}
 \global\long\global\long\global\long\def\bs{\bar{\sigma}}
 \global\long\global\long\global\long\def\mbA{\mathbf{A}}
 \global\long\global\long\global\long\def\mbB{\mathbf{B}}
 \global\long\global\long\global\long\def\mbC{\mathbf{C}}
 \global\long\global\long\global\long\def\mbD{\mathbf{D}}
 \global\long\global\long\global\long\def\mbE{\mathbf{E}}
 \global\long\global\long\global\long\def\mbF{\mathbf{F}}
 \global\long\global\long\global\long\def\mbG{\mathbf{G}}
 \global\long\global\long\global\long\def\mbH{\mathbf{H}}
 \global\long\global\long\global\long\def\mbI{\mathbf{I}}
 \global\long\global\long\global\long\def\mbJ{\mathbf{J}}
 \global\long\global\long\global\long\def\mbK{\mathbf{K}}
 \global\long\global\long\global\long\def\mbL{\mathbf{L}}
 \global\long\global\long\global\long\def\mbM{\mathbf{M}}
 \global\long\global\long\global\long\def\mbN{\mathbf{N}}
 \global\long\global\long\global\long\def\mbO{\mathbf{O}}
 \global\long\global\long\global\long\def\mbP{\mathbf{P}}
 \global\long\global\long\global\long\def\mbQ{\mathbf{Q}}
 \global\long\global\long\global\long\def\mbR{\mathbf{R}}
 \global\long\global\long\global\long\def\mbS{\mathbf{S}}
 \global\long\global\long\global\long\def\mbT{\mathbf{T}}
 \global\long\global\long\global\long\def\mbU{\mathbf{U}}
 \global\long\global\long\global\long\def\mbV{\mathbf{V}}
 \global\long\global\long\global\long\def\mbW{\mathbf{W}}
 \global\long\global\long\global\long\def\mbX{\mathbf{X}}
 \global\long\global\long\global\long\def\mbY{\mathbf{Y}}
 \global\long\global\long\global\long\def\mbZ{\mathbf{Z}}
 \global\long\global\long\global\long\def\mba{\mathbf{a}}
 \global\long\global\long\global\long\def\mbb{\mathbf{b}}
 \global\long\global\long\global\long\def\mbc{\mathbf{c}}
 \global\long\global\long\global\long\def\mbd{\mathbf{d}}
 \global\long\global\long\global\long\def\mbe{\mathbf{e}}
 \global\long\global\long\global\long\def\mbf{\mathbf{f}}
 \global\long\global\long\global\long\def\mbg{\mathbf{g}}
 \global\long\global\long\global\long\def\mbh{\mathbf{h}}
 \global\long\global\long\global\long\def\mbi{\mathbf{i}}
 \global\long\global\long\global\long\def\mbj{\mathbf{j}}
 \global\long\global\long\global\long\def\mbk{\mathbf{k}}
 \global\long\global\long\global\long\def\mbl{\mathbf{l}}
 \global\long\global\long\global\long\def\mbm{\mathbf{m}}
 \global\long\global\long\global\long\def\mbn{\mathbf{n}}
 \global\long\global\long\global\long\def\mbo{\mathbf{o}}
 \global\long\global\long\global\long\def\mbp{\mathbf{p}}
 \global\long\global\long\global\long\def\mbq{\mathbf{q}}
 \global\long\global\long\global\long\def\mbr{\mathbf{r}}
 \global\long\global\long\global\long\def\mbs{\mathbf{s}}
 \global\long\global\long\global\long\def\mbt{\mathbf{t}}
 \global\long\global\long\global\long\def\mbu{\mathbf{u}}
 \global\long\global\long\global\long\def\mbv{\mathbf{v}}
 \global\long\global\long\global\long\def\mbtv{\mathbf{\tilde{v}}}
 \global\long\global\long\global\long\def\mbw{\mathbf{w}}
 \global\long\global\long\global\long\def\mbx{{\mathbf{x}}}
 \global\long\global\long\global\long\def\mby{\mathbf{y}}
 \global\long\global\long\global\long\def\mbty{\mathbf{\tilde{y}}}
 \global\long\global\long\global\long\def\mbz{\mathbf{z}}
 \global\long\global\long\global\long\def\mbzero{\mathbf{0}}
 \global\long\global\long\global\long\def\mbalpha{{\boldsymbol{\alpha}}}
 \global\long\global\long\global\long\def\mbdelta{{\boldsymbol{\delta}}}
 \global\long\global\long\global\long\def\mbepsilon{{\boldsymbol{\epsilon}}}
 \global\long\global\long\global\long\def\mbmu{{\boldsymbol{\mu}}}
 \global\long\global\long\global\long\def\mbnu{{\boldsymbol{\nu}}}
 \global\long\global\long\global\long\def\mbPi{{\boldsymbol{\Pi}}}
 \global\long\global\long\global\long\def\mbpi{{\boldsymbol{\pi}}}
 \global\long\global\long\global\long\def\mbtheta{{\boldsymbol{\theta}}}
 \global\long\global\long\global\long\def\mbphi{{\boldsymbol{\phi}}}

\global\long\global\long\global\long\def\Definition#1#2{\vspace{0.1in} \noindent{\begin{minipage}{\linewidth}\underline{{\bf Def'n:}} {\it #1}\end{minipage}}\vspace{0.1in} }
 \global\long\global\long\global\long\def\Lemma#1#2{\vspace{0.1in}}
 \global\long\global\long\global\long\def\Theorem#1#2{\vspace{0.1in} \noindent\begin{minipage}{\linewidth}\underline{{\bf Theorem:}} {#1} \end{minipage}\vspace{0.1in} }
 \global\long\global\long\global\long\def\StartOf#1{{\noindent\rule{\linewidth}{1pt}}}
 \global\long\global\long\global\long\def\Today#1{{\noindent\large Today: #1}}
 \global\long\global\long\global\long\def\Example#1{{\vspace{0.15in} \noindent{\bf Example: #1}}}
 \global\long\global\long\global\long\def\Note#1{{\vspace{0.05in} \noindent{\bf Note:} #1}}
 \global\long\global\long\global\long\def\pdfarray#1#2{{ \left\{  \begin{array}{ll}
 {#1},  &  {#2} \\
0,  &  o.w. \end{array}\right. }}
 \global\long\global\long\global\long\def\pdfarrays#1#2#3#4{{ \left\{  \begin{array}{ll}
 {#1},  &  {#2} \\
{#3},  &  {#4} \\
0,  &  o.w. \end{array}\right. }}
 \global\long\global\long\global\long\def\twooptions#1#2#3#4{{ \left\{  \begin{array}{ll}
 {#1},  &  {#2} \\
{#3},  &  {#4} \end{array}\right. }}
 \global\long\global\long\global\long\def\threeoptions#1#2#3#4#5#6{{ \left\{  \begin{array}{ll}
 {#1},  &  {#2} \\
{#3},  &  {#4} \\
{#5},  &  {#6} \end{array}\right. }}
 \global\long\global\long\global\long\def\CDFarrays#1#2#3#4{{ \left\{  \begin{array}{ll}
 0,  &  {#3} < {#2} \\
{#1},  &  {#2} \le{#3} < {#4} \\
1,  &  {#3} \ge{#4} \end{array}\right. }}
 \global\long\global\long\global\long\def\CDFarray#1#2#3{{ \left\{  \begin{array}{ll}
 0,  &  {#3} < {#2} \\
{#1},  &  {#3} \ge{#2} \end{array}\right. }}
 \global\long\global\long\global\long\def\CCDFarrays#1#2#3#4{{ \left\{  \begin{array}{ll}
 1,  &  {#3} < {#2} \\
{#1},  &  {#2} \le{#3} < {#4} \\
0,  &  {#3} \ge{#4} \end{array}\right. }}
 \global\long\global\long\global\long\def\CCDFarray#1#2#3{{ \left\{  \begin{array}{ll}
 1,  &  {#3} < {#2} \\
{#1},  &  {#3} \ge{#2} \end{array}\right. }}
 \global\long\global\long\global\long\def\Vector#1{\left[ \begin{array}{c}
 #1 \end{array} \right]}
 \global\long\global\long\global\long\def\MatTwoCols#1{\left[ \begin{array}{cc}
 #1 \end{array} \right]}
 \global\long\global\long\global\long\def\MatThreeCols#1{\left[ \begin{array}{ccc}
 #1 \end{array} \right]}
 \global\long\global\long\global\long\def\PP{\vspace{0.11in}\noindent}
 \global\long\global\long\global\long\def\Fourier#1{{\mathfrak{F}\left\{  {#1}\right\}  } }
 \global\long\global\long\global\long\def\IFourier#1{{\mathfrak{F}^{-1}\left\{  {#1}\right\}  } }
 \global\long\global\long\global\long\def\DTFT#1{{\mathrm{DTFT}\left\{  {#1}\right\}  } }
 \global\long\global\long\global\long\def\IDTFT#1{{\mathrm{DTFT}^{-1}\left\{  {#1}\right\}  } }
 \global\long\global\long\global\long\def\rect{\mbox{rect}}
 \global\long\global\long\global\long\def\sinc{\mbox{sinc}}
 \global\long\global\long\global\long\def\Summary#1{{\vspace{0.15in} \noindent{\bf Summary of today's lecture}}: #1}
 \global\long\global\long\global\long\def\rank{\mbox{rank}}
 \global\long\global\long\global\long\def\atantwo#1#2{{\mbox{atan2} \left({#1}, {#2} \right) }}
 \global\long\global\long\global\long\def\erf{\mbox{erf}}
 \global\long\global\long\global\long\def\erfc{\mbox{erfc}}

\global\long\global\long\global\long\def\En{\mathcal{E}}
 \global\long\global\long\global\long\def\Ebno{\frac{\mathcal{E}_{b}}{N_{0}}}
 \global\long\global\long\global\long\def\Question{{\vspace{0.15in} \noindent{\bf Question: }}}
 \global\long\global\long\global\long\def\dB{\mbox{(dB)}}
 \global\long\global\long\global\long\def\dBm{\mbox{(dBm)}}
 \global\long\global\long\global\long\def\dBW{\mbox{(dBW)}}
 \global\long\global\long\global\long\def\intinfty#1#2{{\int_{#1 = -\infty}^{\infty} #2 d{#1} }}
 \global\long\global\long\global\long\def\decision#1#2{{{#1 \atop >} \atop {< \atop #2}}}
 \global\long\global\long\global\long\def\Q#1{{\mbox{Q}\left( {#1} \right)}}
 \global\long\global\long\global\long\def\Qinv#1{{\mbox{Q}^{-1}\left( {#1} \right)}}
 \global\long\global\long\global\long\def\sgn#1{{\mbox{sgn}\left\{  {#1} \right\}  }}
 \global\long\global\long\global\long\def\entropy#1{H\left[ {#1} \right]}
 \global\long\global\long\global\long\def\stdPM{(P,\mathcal{F}, \Omega)}
 \global\long\global\long\global\long\def\apriori{\emph{a priori} }
 \global\long\global\long\global\long\def\nnn{}
 \global\long\global\long\global\long\def\nn{}
 \global\long\global\long\global\long\def\sifi{$\sigma$-field}
 \global\long\global\long\global\long\def\communicateswith{{\leftrightarrow}}
 \global\long\global\long\global\long\def\MatFourCols#1{\left[ \begin{array}{cccc}
 #1 \end{array} \right]}
 \global\long\global\long\global\long\def\MatFiveCols#1{\left[ \begin{array}{ccccc}
 #1 \end{array} \right]}
 \global\long\global\long\global\long\def\MatSixCols#1{\left[ \begin{array}{cccccc}
 #1 \end{array} \right]}
 \global\long\global\long\global\long\def\MatSevenCols#1{\left[ \begin{array}{ccccccc}
 #1 \end{array} \right]}
 \global\long\global\long\global\long\def\MatEightCols#1{\left[ \begin{array}{cccccccc}
 #1 \end{array} \right]}
 \global\long\global\long\global\long\def\MatNineCols#1{\left[ \begin{array}{ccccccccc}
 #1 \end{array} \right]}
 \global\long\global\long\global\long\def\limin#1{\lim_{{#1}\rightarrow\infty}}

\global\long\global\long\global\long\def\multiplefloor#1#2{{\left\lfloor {#1} \right\rfloor _{#2} }}
 \global\long\global\long\global\long\def\multipleceil#1#2{{\left\lceil {#1} \right\rceil _{#2} }}
 \global\long\global\long\global\long\def\atan{\mbox{atan} }

We investigate an attack to the privacy of the location of people moving in an area covered by a wireless network. People moving in an area covered by one or more wireless networks, affect the way radio signals propagate.  We demonstrate that the presence, location and direction of movement of people not carrying any wireless device can be ``eavesdropped'' by using the channel information of wireless links artificially created by an attacker by deploying sensing devices or \emph{receivers} that can ``hear'' {\em transmitters} such as WiFi access points (APs), composing the legitimate wireless network.  Signals from the transmitters passing through non-metal external walls that allow radio waves to go through, are analogous to light from light bulbs passing through glass windows which an adversary can use to ``see'' where people are in a building.
 Hence, we call this attack on location privacy of people an ``exploiting radio windows'' (ERW) attack.  


\begin{figure}
    \begin{centering}
        \epsfig{figure=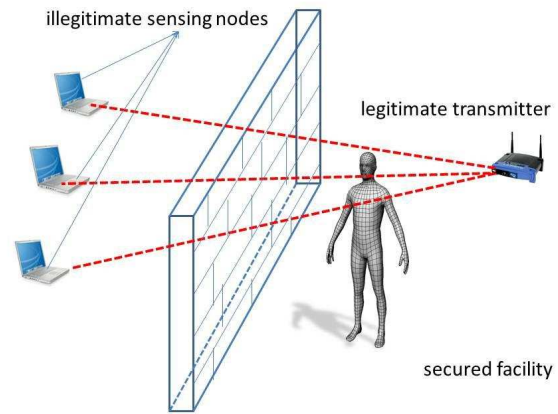,width=\columnwidth}
    \par\end{centering}
    \caption{Exploiting Radio Windows (ERW) attack example.}
    \label{fig:intro}
\end{figure}

Consider a building where security is important (e.g., an embassy) with a concrete exterior wall. One or more wireless networks may have been set up in this building to transfer different types of data, including voice and video.  We can expect these networks to implement advanced data security protocols to prevent eavesdropping of data. However, an attacker can still deploy receivers outside the wall of the building 
to measure different parameters of the received radio signals. By measuring the channel state information (CSI) or received signal strength (RSS), for example, of the links from the transmitters inside the building to the receivers deployed, the attacker can monitor the movements of people and objects inside the building in the area behind the wall in Figure~\ref{fig:intro}. The information about people's movements can be put to malicious use including planning a physical attack on the personnel inside the building. On the contrary, law enforcement personnel can apply similar techniques in the case of a hostage situation to track activity inside a large building and plan their operation accordingly. 

In this paper, we design and implement the ERW attack methodology for through wall people localization. Our methodology relies on reliably detecting when people cross the link lines between the legitimate transmitters and the attack receivers. We first develop a majority-vote based detection algorithm that reliably detects line of sight (LOS) crossing between the legitimate transmitter and the attack receivers by comparing short-term variances in link channel information with its long-term counterpart. We also develop a method to estimate the direction of movement of a person from the sequence of link lines crossed during a short time interval. Next, we implement our methodology on WiFi and ZigBee nodes and experimentally evaluate the ERW attack by monitoring people movements through walls in two real-world settings -- a hallway of a university building separated from the outside by a one-foot thick concrete wall, and a residential house. When we use two WiFi 802.11n nodes with normal antenna separation, or two groups of ZigBee nodes as attack receivers, we find that our methods achieve close to $100\%$ accuracy in detecting line crossings and the direction of movement. We also find that our methods achieve $90-100\%$ accuracy when we use a single 802.11n attack receiver.

To protect the privacy of the location information from the ERW attack, the owner of the legitimate network may choose to implement a countermeasure in which the transmitters vary their transmit power during successive transmissions. The artificial transmit power changes can be either random or follow a pre-defined profile replicating the typical channel variations introduced when a person crosses a link line. This countermeasure is expected to introduce additional variability in the received signal measured by the attack receivers which can be wrongly interpreted by the attacker as caused by moving people or objects crossing the link lines. In this paper, we demonstrate that an attacker who can measure a sufficient number of links can accurately estimate the artificial transmit power change, compensate for it, and ultimately locate people and monitor their movements. We base our compensation strategy on the following intuition: An artificial transmit power change at a transmitter will impact all the links between the transmitter and the attack receivers, whereas genuine power changes due to human movement are likely to impact only some of the links. 

The ERW attack described in this paper is significantly different than device-free localization\footnote{in which people who are not carrying any radio transmitters are located by a static deployed network.} (DFL) in that the ERW attack is practical for large buildings, is stealthy because no transmitters are deployed by the attacker, and is immune from jamming.  DFL systems such as the ones in \cite{zhang2007rf,Wilson_RTI,kanso09b,Wilson_RTI,chen11sequential,kaltiokallio2011,martin2011modelling,Grandma2012,VRTI} require dozens of radio transceivers deployed throughout or on many sides of the target area.  Further, through-building DFL systems such as \cite{VRTI,zheng2012through} assume the transmitted signal penetrates through two external walls and any internal walls in between, and as such have been tested only in buildings of small (18 - 42 m$^2$) size.  In this paper, we show access to one side is sufficient for an ERW attack, and it requires a signal from inside a building to penetrate only one external wall.  Other fingerprint-based DFL systems \cite{Nuzzer_2012,viani2010electromagnetic,kosba2012rasid,xu2012improving} require collection of training data with a person in each possible location in the environment.  In our ERW attack, we do not assume that an attacker has prior access to the inside of the building to be able to perform such data collection.  Further, to perform DFL, an attacker must deploy some nodes which transmit, exposing them to being detected and located by RF source localization, while an ERW attack is stealthier in that purely passive receivers are deployed by an attacker.  Finally, DFL systems' signals could be interfered with by a powerful jammer.  In the method in this paper, any transmitter in the building, including a jammer, could be used as a source for ERW.

The remainder of the paper is organized as follows.
In Section \ref{sec:adversary}, we describe the adversary model.
In Section \ref{sec:method}, we formulate the methods used to detect link line crossings and estimate changes in transmit power. We also describe the method used to determine the direction of motion of the person.
The experimental setup is presented in Section \ref{sec:experiment}. 
In Section \ref{sec:result}, we present the results of our experiments. 
Section \ref{sec:related_work} discusses the previous research in the area of location privacy attacks in wireless networks.
Conclusions and directions for future work are given in Section \ref{sec:conclusion_and_future}.

\label{sec:intro}

\section{Adversary Model}
\label{sec:adversary}

We make the following assumptions about the attacker\footnote{In this paper, we use the term attacker for anyone, whether malicious or genuine, who is trying to localize humans.}:

\begin{itemize}
\item The attacker is able to deploy multiple wireless sensing devices within the transmission range of the legitimate transmitter(s) outside the area being monitored. The attacker is able to measure the physical layer information (RSS and/or CSI) of the links between the transmitter(s) and the attack receivers.  
\item The attacker does not have access to the content of the packets transmitted by the legitimate network nodes.
\item The attacker does not deploy any transmitters, nor does it have any control over the legitimate transmitters. However, it requires the legitimate transmitters to transmit packets frequently to allow it perform the line crossing detections.
\item The attacker does not make any assumption regarding the transmit power profile of the transmitters.
\item The attacker nodes do not associate or interfere in any manner with the transmissions of the legitimate transmitter(s).
\item The attacker may not know the precise
location of the transmitters or the arrangement of their antennas. However, we do assume that a transmitter is located well inside the perimeters of buildings for network coverage reasons ensuring that
they do not lie between the people (being localized) and the attack receivers.
\end{itemize}

\section{Methodology}
\label{sec:method}

In this section, we first develop a methodology to detect line crossings based
on a majority vote for WiFi 802.11n receivers. We also develop a method that uses a sequence of line crossings to determine the direction of human movement. Next, we present our approaches for estimating transmit power change and its compensation, when the transmit power is artificially changed by the owner of the wireless transmitters, inside the a secure building, with the hope of preserving location privacy. Last, we show how we adapt our
methodology for IEEE $802.15.4$ ZigBee attack receivers.

\subsection{Line Crossing Detection}
\label{sec:linewifi}
Many modern WiFi
networks use the 802.11n standard, in which transceivers are
equipped with multiple antennas in order to leverage the spatial diversity of
the wireless channel. While these multiple-input multiple-output (MIMO) systems provide high data rates, they also provide a rich source
of channel information to an adversary interested in localizing people inside a building.

The 802.11n wireless standard uses the well-known orthogonal frequency-division
multiplexing (OFDM) modulation scheme, which encodes and transmits data across
multiple subcarriers for each transmitter-receiver antenna pair. When an 802.11n receiver receives a packet, it estimates the effect of the
wireless channel on each MIMO OFDM subcarrier for the purpose of channel
equalization. Since this channel state information (CSI), represented as a
complex gain for each subcarrier, is measured during the unencrypted preamble of
each WiFi packet, an adversary without legitimate access to data on the network
can still measure the CSI for every packet.

We apply a windowed variance method for detecting abrupt changes in the CSI for
a WiFi link. Let $H_{j,k}(n)$ be the magnitude of the signal strength for the $j$th transmitter-receiver
antenna pair and the $k$th OFDM subcarrier for the $n$th packet. We define the 
windowed variance measurement at packet $n$ as follows. Let
\begin{equation}
  \bar{H}^w_{j,k}(n) = \frac{1}{w}\sum_{i=n-w+1}^{n}H_{j,k}(i),
\end{equation}
\begin{equation}
  \label{eq:winVar}
  v_{j,k}^w(n) = \frac{1}{w-1}\sum_{i=n-w+1}^{n}(H_{j,k}(i) - \bar{H}^w_{j,k}) ^2,
\end{equation}
and 
\begin{equation}
  \label{eq:winStd}
  s_{j,k}^w(n) = \sqrt{v_{j,k}^w(n)},
\end{equation}
where, $w$ is the number of previous CSI samples in the window. We define the
subcarrier-average variance for packet $n$ for a given antenna pair $j$ as
\begin{equation}
  \label{eq:linkVar}
  V_j^w(n) = \frac{1}{N}\sum_{k}v_{j,k}^w(n),
\end{equation}
where $N$ is the number of subcarriers. We define the subcarrier-average
standard deviation for packet $n$ as
\begin{equation}
  \label{eq:linkStd}
  S_j^w(n) = \frac{1}{N}\sum_{k}s_{j,k}^w(n).
\end{equation}
The quantities (\ref{eq:linkVar}) and (\ref{eq:linkStd}) represent the average
CSI variance and standard deviation across all subcarriers for antenna pair $j$
at packet $n$ for a time window which includes the past $w$ packets. We track
both (\ref{eq:linkVar}) and (\ref{eq:linkStd}) over a short-term time window $w_s$,
and a long-term time window $w_l$, allowing us to compare the short-term and
long-term statistics of the WiFi link and detect line crossings.

A line crossing is detected for antenna pair $j$ when 
\begin{equation}
  \label{eq:detect}
  \sum_{n \in D}V_j^{w_s}(n) - V_j^{w_l}(n) > \gamma(n),
\end{equation}
where $D$ is the most recent contiguous set of packets for which $V_j^{w_s}(n) -
V_j^{w_l}(n) > 0$ and the threshold $\gamma(n)$ is defined as
\begin{equation}
  \label{eq:threshold}
  \gamma(n) = V_j^{w_l}(n) + C S_j^{w_l}(n).
\end{equation}
The constant $C$ is included to allow the user to adjust the trade-off between
false alarms and missed detections. 

In the case where there are more than two antenna pairs, we take the majority 
vote between antenna pairs over the short-term window to decide if a line 
crossing has occurred. More specifically, when a receiver antenna detects a line crossing, we count 
the line crossing detections for all the receiver antennas over the short-term window, $w_s$. For a $3\times3$ MIMO transmitter and receiver, this would mean computing a majority vote over nine measurements. When the majority of the receiver antennas detect a line crossing within $w_s$, we infer that a person has crossed the link line between the transmitter and the receiver. We will 
show that this majority vote method improves the performance of our detector by 
decreasing false alarms and missed detections. We decrease the false alarm rate
further by combining temporally close detections together. More specifically, 
if we detect a line crossing at time $t_1$ for a transmitter-receiver pair using 
the majority vote, we do not consider any other line crossing detected in the 
time interval $(t_1,t_1+\Delta]$ for the same transmitter-receiver pair, i.e., all
line crossings detected in the interval $[t_1,t_1+\Delta]$ are considered as a 
single line crossing for a transmitter-receiver pair. 

We note that our window-based variance method differs from the method
presented in \cite{Youssef2007, Nuzzer_2012}. In \cite{Youssef2007, Nuzzer_2012}, Youssef et al. compare recent window-based variance measurements of RSSI at multiple WiFi links to measurements made during a static calibration period when nobody is moving in the area of interest. If a certain number of WiFi links within the area of interest detect motion within a certain time interval, a motion event is detected in the area of interest. Our attacker does not know if and/or when people are moving inside of the building, and therefore cannot create calibration measurements based on a static environment. Instead, we compare a short-term window variance to a long-term windowed variance. The long-term window allows us to capture the behavior of the wireless links when the majority of measurements are likely made while there is nobody crossing the link line. Additionally, in the case of 802.11n, we exploit the effect that line crossings have on each OFDM subcarrier and MIMO antenna pair. 

\subsection{Determining Direction of Motion}
\label{sec:directwifi}

If the adversary measures the CSI at multiple receivers, or if a single receiver includes
multiple antennas as is the case with 802.11n, it is also possible to infer the
direction that a person is walking when line crossings are detected. The
direction of motion is inferred from the time differences between the line
crossing detections at each receiver, in the case of multiple receivers, or at each
transmitter-receiver antenna pair, when the receivers include multiple antennas.

\begin{figure*}
  \begin{center}
    \mbox{
        \subfigure[\quad]{\epsfig{figure=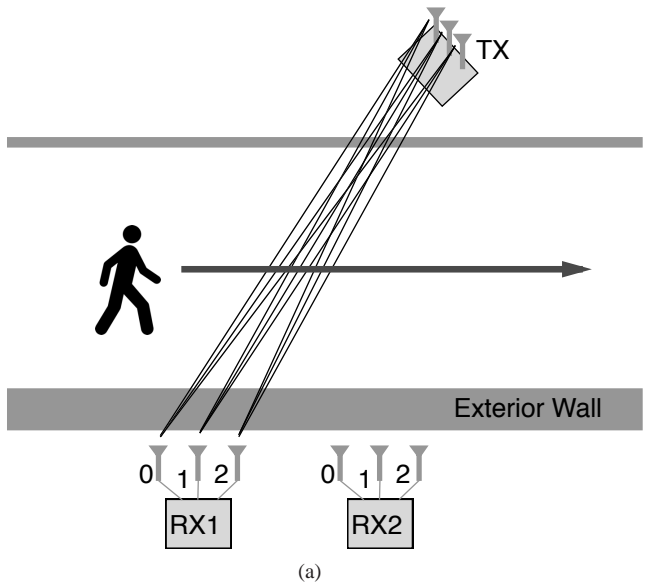,width=\columnwidth}} \quad
        \subfigure[\quad]{\epsfig{figure=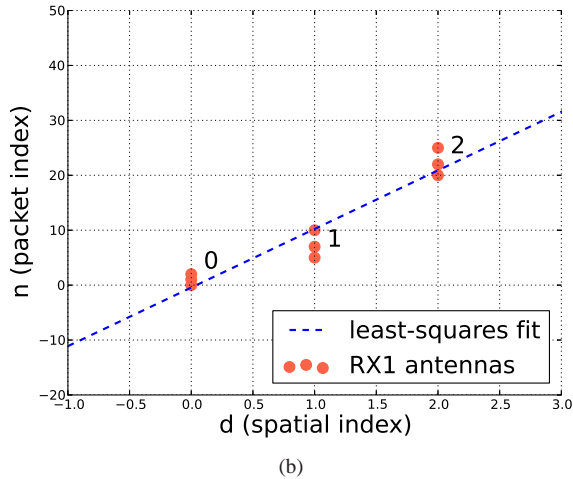,width=\columnwidth}}
        }
    \caption{(a) Line crossing detection diagram. The attack receiver(s) 
     measure channel state information from the legitimate transmitter. The MIMO antenna array at the receiver(s) allows the adversary to count line crossings and determine
    direction of motion. (b) Direction of motion is determined by fitting a line
    to the points created by the spatial indexes of the antennas which detect a
    line crossing and the corresponding packet indexes of the detections. The sign
    of the slope of the fitted line indicates the direction of motion.}
  \label{fig:diagram}
  \end{center}
\end{figure*}

Consider the scenario where the attacker arranges the MIMO antenna array of
an 802.11n receiver such that the antennas are roughly parallel to a hallway
as shown in Figure~\ref{fig:diagram}(a). The spatial order of the antennas with
reference to the hallway is known, and each transmitter-receiver antenna is given an index according to its spatial order.  Based
on the adversary model assumption that a transmitter is located well
inside the perimeter, the attacker, even without knowing the precise
location of the transmitter or the arrangement of its antennas,
may treat the antennas of the wireless transmitter as if they are co-located and still
achieve reliable results.

In the single WiFi receiver case, if a link crossing is detected by majority vote for a
given short-term window, we find the line that best fits the set of points
$\{(d_j,n_j):j \in P\}$, where $d_j$ is the spatial index of antenna 
pair $j$ representing it's location relative to the other links, $n_j$ is the packet index indicating when a detection occurred at antenna pair $j$ according to (\ref{eq:detect}), and $P$ is the set 
of antenna pairs ending at the WiFi receiver which detected a line crossing during the 
short-term window.  The sign of the slope of this line indicates the direction 
of motion. Figure \ref{fig:diagram} shows an example which uses CSI measurements 
from three antennas at the WiFi transmitter and three antennas at WiFi RX1 (9 antenna pairs). In 
the case of two single-input single-output (SISO) WiFi receivers, a similar method may be 
applied, but the two spatial and packet indexes directly determine the line and its slope.

\subsection{Compensation of Transmit Power Change}
\label{sec:txchange}
In this subsection, we propose a methodology to detect artificial transmit power changes (if any) and compensate for the same. The signal strength  
for the $j$th transmitter-receiver antenna pair and the $k$th OFDM subcarrier 
for packet $n$ is given by
\begin{equation}
  {H}_{j,k}(n) = T_x(n) + G_t + G_r - L_{j,k}(n) + \Psi_{j,k}(n),\label{eq:RSS_meast}
\end{equation}
where $T_x(n)$ is the transmit power of the transmitter at time $n$, $G_t$ and 
$G_r$ are the transmitter and receiver antenna gains, respectively, $L_{j,k}(n)$ is 
the path loss, and $\Psi_{j,k}(i)$ is a noise term.  The path loss includes 
all environmentally-dependent terms, including large-scale loss, shadowing, and 
small-scale fading.  The noise term includes thermal noise, quantization noise, 
and other measurement noise at the attacker receiver.  

The attacker cannot depend on knowing the transmit power or antenna gains.  
Instead, the attacker relies on the difference between the subcarrier signal strength for the packet $n$ and the reference packet ($n=0$) (the attacker may update
the reference packet periodically to account for changes in the environment). This difference in subcarrier signal strength is given by
\begin{equation}\label{eq:RSS_difference_defn}
 h_{j,k}(n) \triangleq H_{j,k}(n) - H_{j,k}(0).
\end{equation}
From (\ref{eq:RSS_meast}), we see that
\begin{equation}\label{eq:attack}
 h_{j,k}(n) = t_x(n)- l_{j,k}(n) + \psi_{j,k}(n),
\end{equation}
where
\begin{eqnarray}
t_x(n) &=& T_x(n) - T_x(0), \nonumber \\
l_{j,k}(n) &=& L_{j,k}(n)  - L_{j,k}(0), \nonumber \\
\psi_{j,k}(n) &=& \Psi_{j,k}(n) - \Psi_{j,k}(0). \nonumber
\end{eqnarray}
The subcarrier signal strength difference $h_{j,k}(n)$ above, contains transmit power changes and 
channel-induced changes between the $n$th packet and the reference packet, in addition to noise.

The ideal situation from the attacker's perspective would be that there is no artificial change in transmit powers, and that 
$t_x(n)=0$ for all $j$ and $k$. In this ideal situation, the subcarrier signal strength difference below is solely due to changes in the channel.
\begin{equation}\label{eq:rssnoattack}
 h_{j,k}(n) = - l_{j,k}(n) + \psi_{j,k}(n).
\end{equation}
Furthermore, people crossing the line between the transmitter and receiver antennas 
typically cause a path loss change more significant than noise, and thus the
 $h_{j,k}$ signal allows direct inference of people's motion.  However, when 
the transmitter artificially changes its transmit power, from (\ref{eq:attack}), we cannot directly attribute a large magnitude of $h_{j,k}$ to environmental changes. In particular, if the magnitude of transmit power changes is high enough, the magnitude of $h_{j,k}(n)$ will be predominantly due to because of transmit power changes at the transmitter. A transmitter could thus presumably preserve location privacy by changing its transmit power frequently.

We now propose a method that a smart attacker can use to estimate and remove the artificial power changes and accurately detect
line crossings. In our method, the attacker estimates the artificial transmit power change amplitude by correlating measurements across all antenna pairs and all subcarriers, and removes the effect of transmit power changes from the received signal strength measurements. We propose to use the median of $h_{k,j}(n)$ for all available transmitter-receiver
antenna pairs and corresponding subcarriers, as an estimator of the artificial transmit power change, as shown in the equation below: 
\begin{equation}
\hat{t}_x(n) = \mbox{median}\left\{ h_{j,k}(n) \forall{j,k} \right\}. \label{eq:txest}
\end{equation}
Our choice of this estimator is based on the following observations.
First, we observe that $t_x(n)$ appears in the equation for $h_{k,j}(n)$ for all $j$ and $k$. This is because, any change in transmit power affects measurements across all transmitter-receiver
antenna pairs and corresponding subcarriers simultaneously. Moreover, $t_x(n)$ is linearly related to $l_{j,k}$ and $h_{j,k}$.  
We also know that the change in the path loss $l_{j,k}$ is just as likely to be positive as negative. Furthermore, any change due to human movement will not affect all the links simultaneously.

In the absence of an artificial transmit power change,  $\hat{t}_x(n)$  is likely to be close to zero, i.e., our estimator does not require us to detect whether or not there is an artificial transmit power change for packet $n$.  

The compensated signal strength for packet $n$, which we denote $\hat{H}_{j,k}(n)$, is given by
\begin{equation} \label{eq:compensated_RSS}
\hat{H}_{j,k}(n) = H_{j,k}(n) - \hat{t}_x(n).
\end{equation}
Although the reference packet was sent with unknown 
transmit power $T_x(0)$, for $n>0$, we consider $T_x(n)$ to be the relative dB shift in transmit power compared to $T_x(0)$. $\hat{H}_{j,k}(n)$ essentially, is an estimate of the subcarrier signal strength if there were no transmit power changes between the reference packet and packet $n$.    

It is clear that, any error in the estimation of the transmit power changes amplitude will introduce additional noise in the measurements.  However, the dynamics of the signal are still preserved and an attacker can use any variation in the signal over a short time period in order to notice motion of a person near the link line.  

\subsection{ZigBee Networks}
\label{sec:zigbee_method}
The methodologies described above are also applicable for IEEE $802.15.4$ ZigBee nodes.
However, the ZigBee nodes are generally equipped with a single antenna, so the MIMO setup is not available. Moreover, ZigBee nodes do not use OFDM for communication, so we use channel information
from a single frequency channel (instead of averaging across all subcarriers as in the case of OFDM) to evaluate our methodologies. Furthermore, there is no tool to get the complete CSI at the receiver.
Instead, we rely on the RSS value obtained from the receiver hardware.
Thus, in the case of ZigBee we set $H_{j,k}(n)$ to the RSS value 
measured in decibel units for the $j$th transmitter-receiver antenna pair for packet $n$,
also $k=1, \forall{j}$ as we have measurements from a single channel only.

In order to create spatial diversity we use three closely located ZigBee receivers together to form a group as described in Section \ref{sec:experiment}. We detect line crossings
by applying our majority vote approach on the three links formed between the transmitter and
the three receivers in the group. We detect direction of motion using two groups of receivers
and observing sequence of groups crossed over a short time window. We estimate and compensate
for artificial transmit power changes (if any) by applying the methods described in Section \ref{sec:txchange},
and utilizing the fact that any change in transmit power affects all receivers simultaneously across all groups.

\section{Experiments}
\label{sec:experiment}

In this section, we describe the experimental setup. Section \ref{sec:tools} describes the tools we use to measure the wireless channel, Section \ref{sec:tx_variation} describes the transmit power changes we apply, and Section \ref{sec:deployments} describes two real-world experimental deployments.

\subsection{Tool Description} \label{sec:tools}
We use the following tools to measure the wireless channel and detect line crossings.
\subsubsection{WiFi}
We use laptops with Intel $5300$ NICs that have three-antenna MIMO $802.11n$ radios. We use the CSI Tool \cite{csitool}, that has been built for these radios, to get channel state information from the WiFi transmitter. The CSI tool extracts $802.11n$ channel state 
information for $30$ subcarrier at each antenna pair. Since we use three antennas at each node for communication, for each 
transmitter-receiver pair, we have $3\times3=9$ links each with $30$ subcarrier groups. We use two kinds
of antenna separations - in the normal case (WiFi\_NORM), we place the antennas $~6$ cm apart, in the other case (WiFi\_SEP), we use a larger antenna separation of $~30$ cm. The increased separation is accomplished by connecting the antennas to the Intel 5300 NIC with standard RF cables that are long enough to provide up to $30$ cm separation. 
We program the transmitter to transmit packets at a rate of $~10$ Hz which is similar to beacon frame rates of 
a standard wireless access point. The attack receivers use the CSI Tool to obtain channel state information from the received packets which in turn is used to detect line crossings as described
in Section \ref{sec:linewifi}. 

  \subsubsection{ZigBee}
  For the ZigBee experiments, we use Texas Instrument CC2531 
  USB dongles \cite{tidonglenode}, which are equipped with low-power, IEEE $802.15.4$-compliant radios operating 
  in the $2.4$ GHz ISM band. The transmission frequency in this case is $~12$ Hz. A laptop is used to process the 
  measured data at the attack receivers. There is no tool to obtain the CSI information in the case of ZigBee nodes. Therefore, we use the 
 RSS value (in dBm) measured by the receiver hardware for our analysis, as described in Section \ref{sec:zigbee_method}.
  
  \subsection{Transmit Power Variations} \label{sec:tx_variation}
  We consider three different settings of transmit power variations for our experiments: 
\begin{enumerate}[(a)]
\item TX\_NORMAL: In this case, the transmitters transmit with fixed transmit power and variations in RSS are 
due to person movement and noise only.
\item TX\_LINECROSS: In this case, we simulate the effect of transmit power change by modifying received data 
according to a power profile that replicates typical signal attenuation introduced by a person crossing the link line. 
We randomly select different time points in the measurements to introduce effect of transmit power change.  
\item TX\_RANDOM: Here, we experimentally implement or simulate the scenario where the transmitter may use a different power level for each transmission by randomly selecting from a predefined set of power levels supported by the hardware. For ZigBee nodes, we program the transmitter(s) to change its transmit power at each transmission by randomly selecting one among four pre-defined transmit power levels, \emph{i.e.},
$+4.5$ dBm, $-1.5$ dBm, $-6$ dBm, and $-10$ dBm.  However, we are
unable to program the random power changes in WiFi nodes and hence, we simulate these power changes.
\end{enumerate}  

     While simulating effects of transmit power change we rely on the fact that any change in the transmit power at a time 
     instant is observed across all subcarriers for all transmitter-receiver antenna pairs in case of WiFi and across 
     all receivers in case of ZigBee at the same instant and we change the received signal parameters accordingly. 
     We also add a zero mean Gaussian random variable (with standard deviation $0.67$) to each 
     $H_{j,k}(n)$ measurement, in addition to the the transmit power change $t_x(n)$, to account
     for errors due to environmental noise.

      \begin{figure*}
  \begin{center}
    \mbox{
    \subfigure[\quad University hallway.]{\epsfig{figure=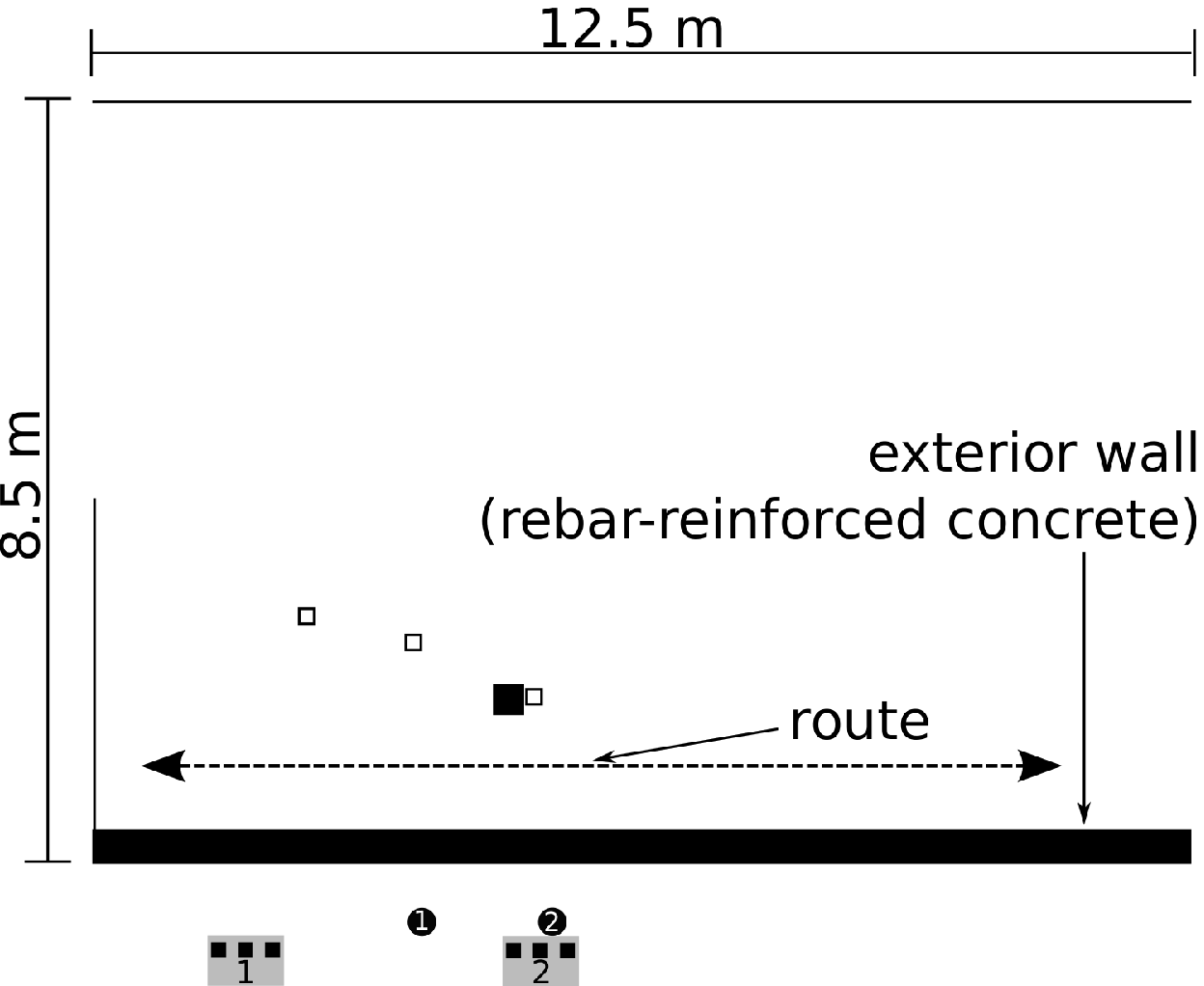,width=0.68\columnwidth} \label{fig:exp_univ_map}
    } 
    \quad\quad\quad\quad\quad\quad
    \subfigure[\quad Residential.]{\epsfig{figure=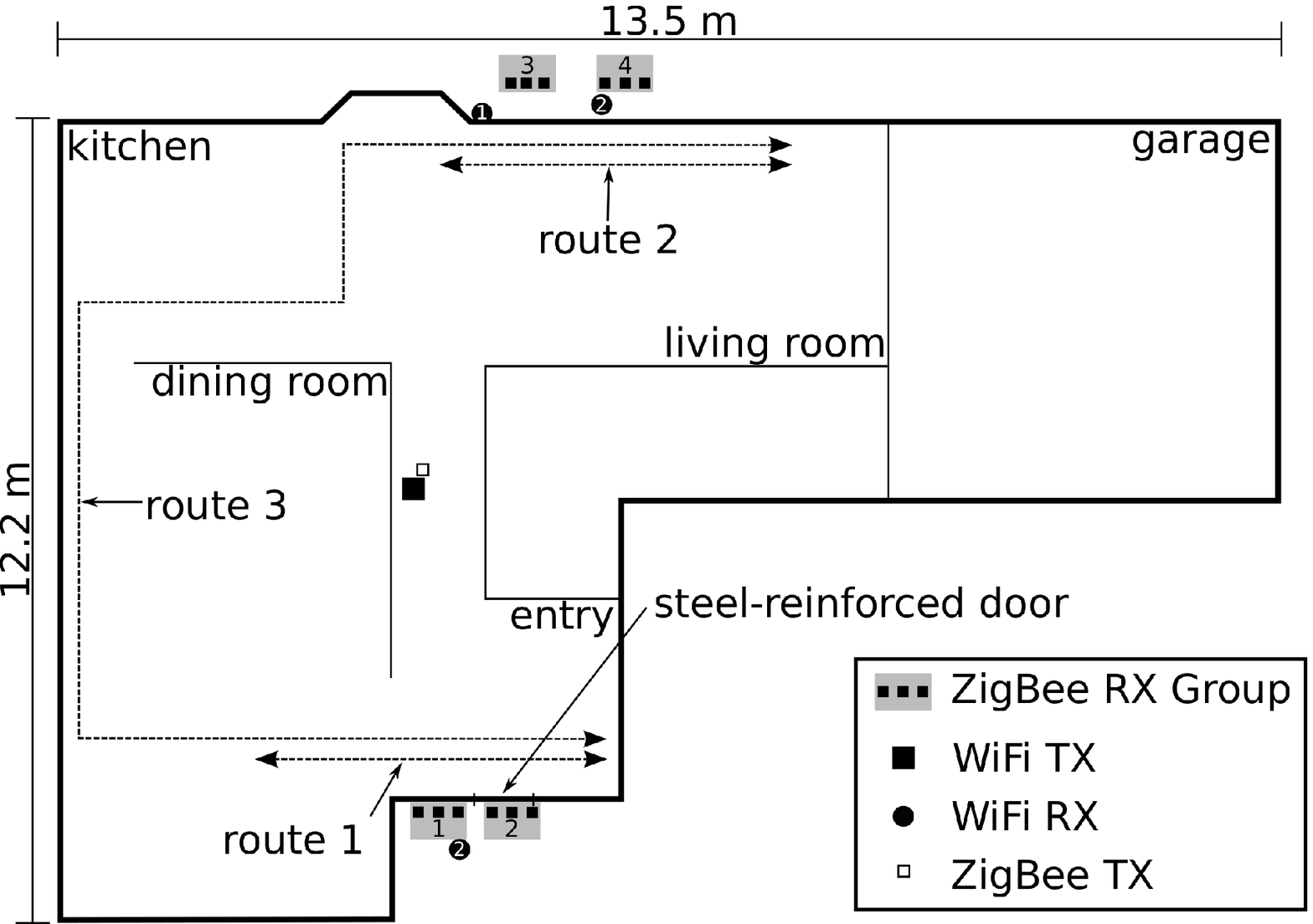,width=\columnwidth} \label{fig:exp_house}}
    }   
  \caption{Network layouts. We show maps of the University Hallway and the Residential House and mark
  the location of the legitimate transmitter(s) and the attack receivers. We also highlight the route(s) followed
  by the walking person.}
  \end{center}
\end{figure*}

\subsection{Experimental Deployments} \label{sec:deployments}
We evaluate our methodologies in two different real world settings. 
\subsubsection{University Hallway}
\begin{figure*}
  \begin{center}
    \mbox{
        \subfigure[\quad Transmitter deployment]{\epsfig{figure=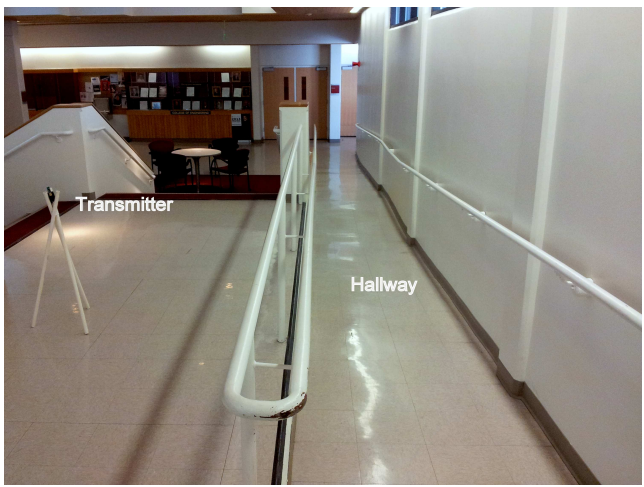,width=\columnwidth,height=2.5in}} \quad
        \subfigure[\quad Attack receiver deployment]{\epsfig{figure=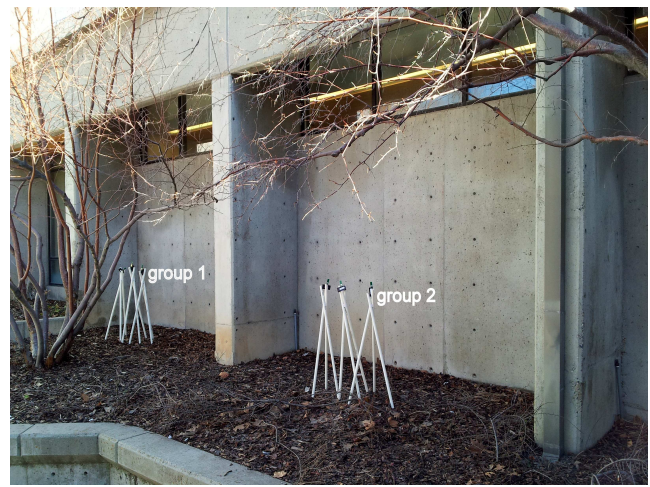,width=\columnwidth,height=2.5in}}
        }
    \caption{Experimental Setup for ZigBee (University Hallway).}
  \label{fig:exp_univ}
  \end{center}
\end{figure*}

  We choose a hallway inside a university building as the area being monitored (Figure \ref{fig:exp_univ}(a)). 
  The hallway is adjacent to a $~30$ cm thick and $~3.5$ m tall rebar-reinforced concrete boundary wall (Figure \ref{fig:exp_univ}(b)). We note that this type 
  of a wall causes significant RF attenuation at WiFi frequencies and represents a worst-case scenario among typical 
  exterior walls for our purposes \cite{stone1997nist}. We place the attack receivers outside the boundary wall parallel 
  to the hallway approximately $1$ m away from the wall.

For the WiFi experiment, we deploy one transmitter inside the building across the hallway, and two attack receivers separated 
by $~3$ m outside the concrete wall (Figure \ref{fig:exp_univ_map}). Similarly, for the ZigBee network, we deploy one transmitter across the hallway and six 
receivers outside the boundary wall. The attack receivers are placed in two groups of three nodes each, with the distance between the 
groups being $~3$ m (Figure \ref{fig:exp_univ_map}). Nodes in the same group are almost $30$ cm apart. We perform both TX\_NORMAL and TX\_RAND experiments with the same ZigBee setup. We also experiment
with three different transmitter locations in case of ZigBee.

  During the experiment, a person walks back and forth along a predefined path (route in Figure \ref{fig:exp_univ_map}) along the corridor between the transmitter and the attack receivers. With the help of a metronome, the person walks at a constant speed of $0.5$ m/s. We collect over $12,000$ data samples for WiFi and over $20,000$ data samples for ZigBee in this experiment. In our evaluation, we use $w_s=4$ s (short time window), $w_l = 40$ s(long term window), and $\Delta=4$ s (Section \ref{sec:linewifi}).

\subsubsection{Residential House}
In this experiment, we monitor two sides of a residential house (Figure \ref{fig:exp_house}) to detect people movement. We perform two sets of experiment with the WiFi nodes. In the first experiment (House 1), we
place the WiFi transmitter in a corridor centrally located inside the house and two WiFi receivers with normal antenna separation (WiFi\_NORM) in the backyard of the house outside the external wall as shown in the Figure \ref{fig:exp_house}. The receivers are placed approximately $1$ m away from each other. For the second experiment (House 2), we use two WiFi receivers with larger antenna separation (WiFi\_SEP) and place one of them in the backyard and the other outside the front entrance. The transmitter is placed in the same position as in experiment House 1.

For the ZigBee network, we place two groups of receivers, each group with three nodes, on either side of the house outside the external walls. As shown in Figure \ref{fig:exp_house}, the ZigBee groups $1$ \& $2$ are placed outside the front entrance, and groups $3$ and $4$ are placed in the backyard, approximately $1$ m away from the walls. Nodes in the same group are almost $30$ cm apart while the inter-group distance on either side being at least $1$ m. The ZigBee transmitter is placed inside the house co-located with the WiFi transmitter. We perform two sets of experiments with the same network settings - in one experiment the ZigBee transmitter transmits with fixed transmit power of $+4.5$ dBm (TX\_NORMAL), in the other experiment the transmitter is programmed to vary its transmit power randomly with each transmission (TX\_RANDOM).

During these residential experiments, a person walks inside the house at normal speed back and forth first near the front entrance of the house (route $1$ in the Figure~\ref{fig:exp_house}), and then in the living room which is near the rear end of the house (shown as route $2$ in the Figure~\ref{fig:exp_house}). Finally, the person makes a few rounds inside the house as shown in route $3$ in the Figure~\ref{fig:exp_house}. We collect over $10,000$ data samples for each set of ZigBee and WiFi experiments. We video record the
line crossings to test the accuracy of our detection method against ground truth. For the residential experiments, we use $w_s=2$ s (short time window), $w_l = 20$ s (long term window) and $\Delta=4$ s  (Section \ref{sec:linewifi}). We use smaller window sizes for detection of line crossings as the person walks at a faster speed as
compared to the University Hallway experiments.



\section{Results}
\label{sec:result}

We evaluate the performance of the ERW attack in terms of false alarm and missed detection rates. False alarm (FA) rates are calculated as the number of line crossings wrongly detected by the system over the number of sample points. Missed detection (MD) rates are calculated as the number of actual line crossings not detected by the system over the total number of actual line crossings. 
\subsection{Detection of Line Crossing}
\label{sec:los}


\begin{table}
\caption{Detection Accuracy (Hallway).}
\begin{tabular}{ |c|c|c|c|c|c|}
 \hline
Hallway  & \multicolumn{2}{c}{Accuracy} \vline & \multicolumn{3}{c}{Error (sec)}\vline\\
\hhline{~-----}
Experiment: & FA\% & MD\% & Min & Max & Mean\\
\hline
WiFi\_NORM & $0$ & $1.92$ & $0.03$ & $2.73$& $0.79$\\
WiFi\_SEP & $0$ & $0$ & $0.27$ & $2.37$& $1.22$\\
ZigBee & $0$ & $1.02$ & $0.27$ & $2.37$& $1.22$\\
\hline
\end{tabular}
\label{table:det_acc}
\end{table}

\begin{table}
\caption{Detection Accuracy (House).}
\begin{tabular}{ |c|c|c|c|c|c|}
 \hline
House   & \multicolumn{2}{c}{Accuracy} \vline & \multicolumn{3}{c}{Error (sec)}\vline\\ \hhline{~-----}
Experiment: & FA\% & MD\% & Min & Max & Mean\\
\hline
WiFi\_NORM & $0.043$ & $5.70$ & $0.29$ & $2.78$& $1.06$\\
WiFi\_SEP & $0.005$ & $4.35$ & $0.03$ & $1.82$& $0.56$\\
ZigBee & $0.004$ & $0.49$ & $0.10$ & $3.55$& $1.63$\\
\hline
\end{tabular}
\label{table:det_acc_house}
\end{table}

\begin{figure}[t]
    \begin{centering}
        \epsfig{figure=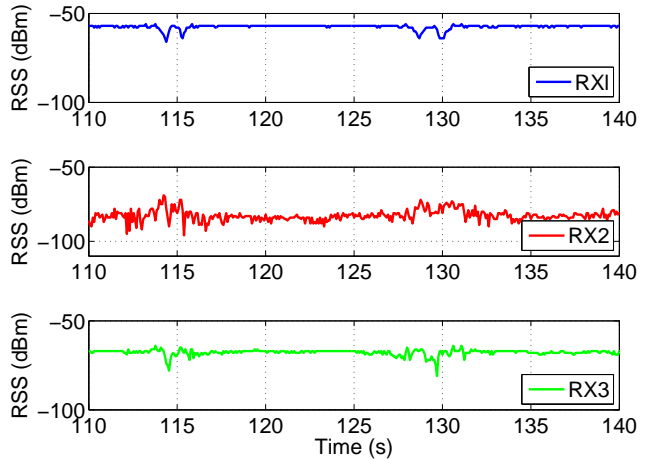,width=\columnwidth}
    \par\end{centering}
    \caption{All links are not equally sensitive to movement - RX1 and RX3 measure high short term 
    variations in link RSS corresponding to person crossings (time intervals [113 s - 116 s] and [128 s - 131 s]). Such a distinct high variation region is not present in link to RX2. }
    \label{fig:link_sens}
\end{figure}

In this section, we present the accuracy of detection of line crossings using 
the methodology as described in Section \ref{sec:linewifi}. 
\subsubsection{University Hallway}
Table \ref{table:det_acc} lists the results obtained in the University Hallway experiment using our majority vote detection. We  achieve almost $100\%$ detection rate with few false alarms and missed detections.
Using a WiFi 802.11n receiver with normal antenna separation, we get zero false alarms
and only $1.92\%$ missed detections. We compare the detected crossing times with those in the recorded video footage of the experiment and find that we can  detect the crossing times with an average error of $0.79$ s, with minimum and maximum errors of $0.03$ s and $2.73$ s respectively. 

We obtain zero false alarms and missed detections
when using a 802.11n WiFi receiver with a large spatial separation between antennas,
the mean error in this case being $1.22$ s. For ZigBee, using a 
group of three closely located receivers, we get a 
$2.66\%$ false alarm rate and a $1.67\%$ missed detection rate in line 
crossing detection with an average error of $1.22$ seconds. We use two groups of receivers and experiment with three different transmitter locations in case of ZigBee. We obtain the above results by averaging over all transmitter location and receiver group pairs. 

Note that while computing the errors as compared to the ground truth, we consider
the line connecting the centroid of transmitter antenna locations
(or the transmitter location in case of ZigBee) 
and the centroid of the receiver antenna locations (or the centroid 
of the receiver locations in the group in case of ZigBee)
as the representative link line.

\subsubsection{Residential House}
We present the detection accuracy of the Residential House experiment in Table \ref{table:det_acc_house}.
We achieve greater than $94\%$ detection accuracy with a $0.043\%$ false alarm rate while using WiFi receivers with normal antenna separation (WiFi\_NORM). With larger antenna separation (WiFi\_SEP) the accuracy is above $95\%$ with a $0.005\%$ false alarm rate. The mean error in detection of line crossings is $1.06s$ in case of WiFi\_NORM, the same being $0.56s$ for WiFi\_SEP.  

For ZigBee, we achieve above $99\%$ accuracy in detection with a false alarm rate of $0.004\%$ only. The mean time-of-crossing estimation in this case is $1.63$ s.
Note that during this experiment, we placed one group of ZigBee nodes (group $2$) directly in
front of the metal-plated entrance door. The packet reception rates for receivers in this group
are much lower than the receivers in the other groups. Also, perhaps due to attenuation through the door, the RSS measurements made by this group are more noisy than those made by the other groups, leading to further degradation in performance. The missed detection rate for
this group is almost $30\%$, about $60$ times more than the average missed detection rate 
of other groups (results presented in Table \ref{table:det_acc_house} are averaged over the other three groups). Thus, we conclude that, although an ERW attack can penetrate concrete and brick walls,
metallic structures in the line of sight path of the radio signals degrades the detection accuracy significantly.

\subsection{Determining Direction of Motion}
\label{sec:movement}
\subsubsection{University Hallway}
In this section, we present the accuracy we achieve in detecting the direction of motion 
in the experiment where we monitor the university hallway. In this experiment, the corridor was crossed by a 
moving person an equal number of times in either direction. We achieve $100\%$ accuracy in detecting direction of movement on either side of 
the corridor while using two WiFi receivers or two groups of ZigBee nodes using the method 
described in Section \ref{sec:directwifi}. 

We also achieve an accuracy as high as $90.38\%$ in detecting direction of motion with only 
a single WiFi 802.11n receiver by increasing the spatial separation of the MIMO antennas. The 
accuracy with a single WiFi receiver with standard antenna 
separation is $59.62\%$, which is slightly better than guessing the direction of motion. 

\subsubsection{Residential House}
For the experiment performed in the residential house we again achieve $100\%$ accuracy in detection while using two WiFi receivers with standard antenna separation (experiment House 1) or two groups of ZigBee nodes on either side of the house. Individual detection accuracy of the two WiFi receivers (with standard antenna separation placed on the same side of the house as in experiment House 1) used are $100\%$ (RX1) and $68\%$ (RX2) respectively. Detection accuracy with spatially separated antennas for these receivers (when they are placed on opposite sides of the house as in experiment House 2) are $96\%$ (RX1) and $52.6\%$  (RX2) respectively. These results differ from the 
University Hallway experiment where we get better accuracy in detecting direction of movement while using large spatial separation between antennas as compared to using normal antenna separation. The degradation in accuracy with antenna separation in Residential House experiment may be due to the fact that during the House 2 experiment, walking speed of the person was about $20\%$ faster as compared to the House 1 experiment with normal antenna separation, hence crossing times for individual antennas overlapped with each other in some cases.

To summarize, our results indicate that an ERW adversary should use two WiFi receivers or two groups of ZigBee nodes at each side in order to detect direction of motion accurately. It is possible to achieve very high accuracy even with a single WiFi receiver in some cases (e.g. RX1 in experiments House 1 and House 2), however the results depend on the environment and need further investigation.

\begin{figure*}[t]
    \begin{centering}
        \epsfig{figure=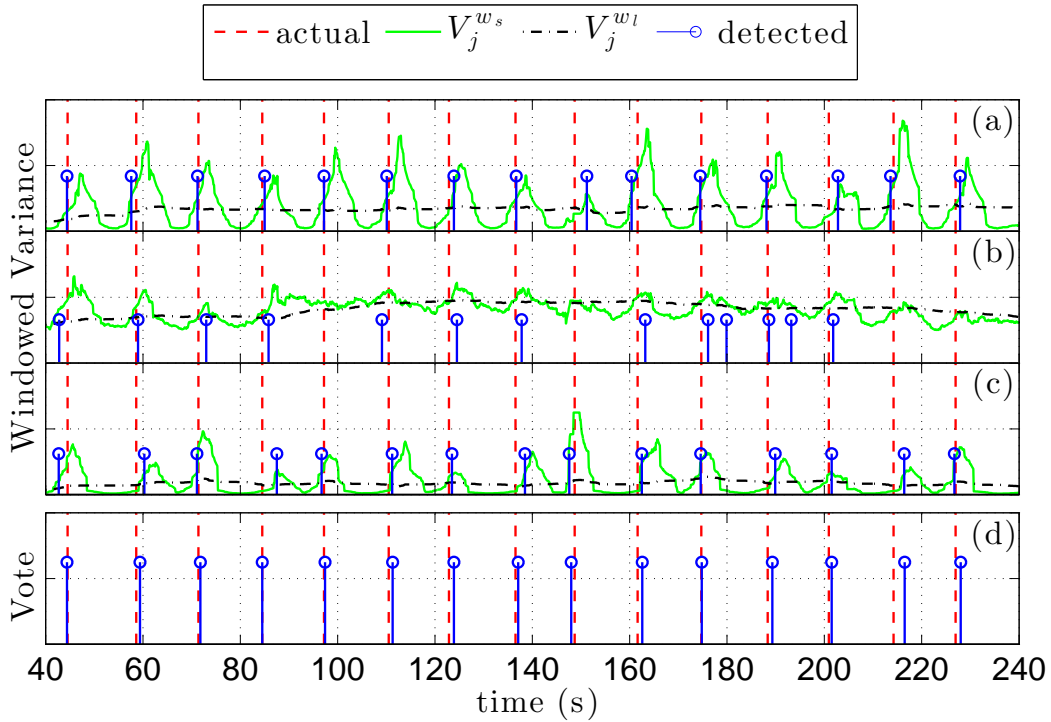,height=4in}
    \par\end{centering}
    \caption{The majority vote over transmitter-receiver antenna pairs reduces
    false alarms and missed detections. (a),(b), and (c) show the results of the
    windowed variance based line crossing detection for three antenna pairs using Wifi. In (d),
    we see that the majority vote eliminates false alarms and missed
    detections.}
    \label{fig:vote}
\end{figure*}

\subsection{Advantages of Majority Vote}
\label{sec:majority}

In this section, we show how our majority vote approach helps overcome inherent
uncertainties in wireless links. All wireless links are not equally sensitive 
to motion occurring in their vicinity and the sensitivity varies with link
fade level along with other factors. For example, Figure \ref{fig:link_sens}, plots the RSS 
for the three ZigBee receivers belonging to group $1$ used in the Residential House 
experiment for a time interval during which the person crossed in front the group two times. 
For RX1 and RX3, the overall RSS variance is very small. When the person crosses the link line, she causes high short term variation of the RSS, as 
can be seen during time intervals [113 s - 116 s] and [128 s - 131 s]. Thus, one can infer link crossing times monitoring for these high short term variations in RSS for these links.
However, the link to RX2 has very low mean RSS value with high variance overall. 
This link does not show clear short term high variance 
region corresponding to actual link line crossings as compared to RX1 or RX3.
Hence, a line crossing detection method that relies only on the link to RX2 will perform poorly.

Since it is not possible for an adversary
to know beforehand whether a link is good or bad for detecting LOS crossings,
he relies on correlation among multiple closely located links and infers a 
line crossing only when majority of these closely located links indicates a
crossing. In our experiments, $3\times3 = 9$ links between the MIMO transmitter-receiver antenna pairs
are considered for majority vote in the WiFi case, and groups of 3 single-antenna receivers in the ZigBee case. Figure \ref{fig:vote} shows one scenario where our 
majority vote algorithm helps get rid of some false alarms and missed detections
due to one bad WiFi link (for clarity we show three out of the nine links) from the University Hallway experiment. 
As can be seen the link in Figure \ref{fig:vote}(b), 
fails to detect a line crossing that occurs around $100$ s, however the 
other two links (Figure \ref{fig:vote}(a) \& Figure \ref{fig:vote}(c)) 
detect the crossing and a majority vote among these three links detects
the crossing at that time (Figure \ref{fig:vote}(d)). Similarly, we see that the link in 
Figure \ref{fig:vote}(b) flags a false alarm at $180$ s but the other
two links do not indicate any crossing. Hence again the majority
vote gets rid of the false alarm at time $180$ s (Figure \ref{fig:vote}(d)),
thereby improving the overall accuracy of the system. 

We summarize our findings as follows - a single wireless link 
 suffices in some cases in detection of line crossings between 
a transmitter and a receiver,  however the results are not always 
reliable due to inherent uncertainties in link sensitivity to 
object movements. 
We can improve accuracy and reliability by correlating detections across
multiple co-located links using a majority vote approach. Our results
confirm that we can get rid of most of the false alarms and missed
detections caused by a bad link by applying the majority vote based
detection method.

\begin{figure*}
  \begin{center}
    \mbox{
        \subfigure[\quad TX\_RANDOM (WiFi)]{\epsfig{figure=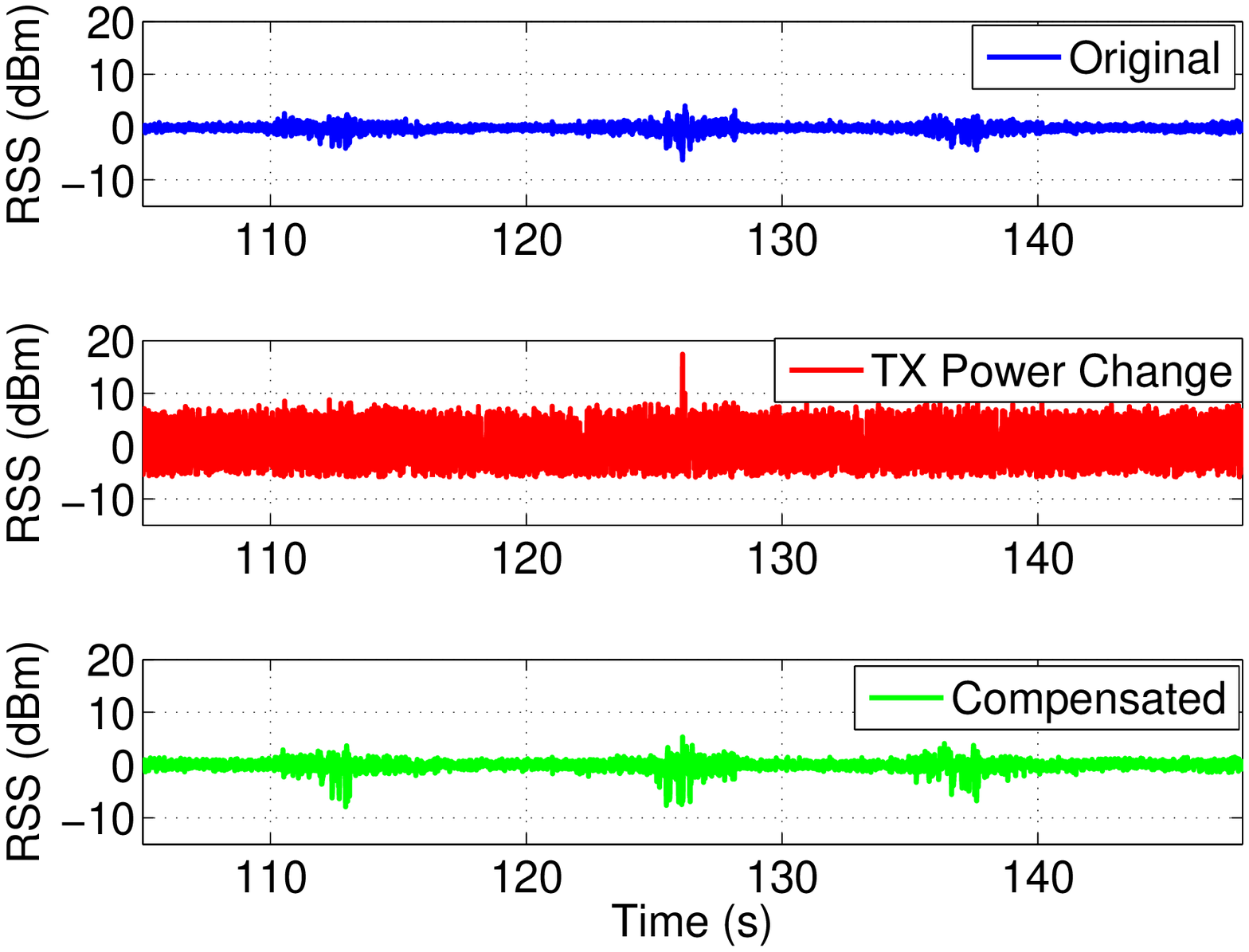,width=\columnwidth}} \quad
        \subfigure[\quad TX\_LINECROSS (ZigBee)]{\epsfig{figure=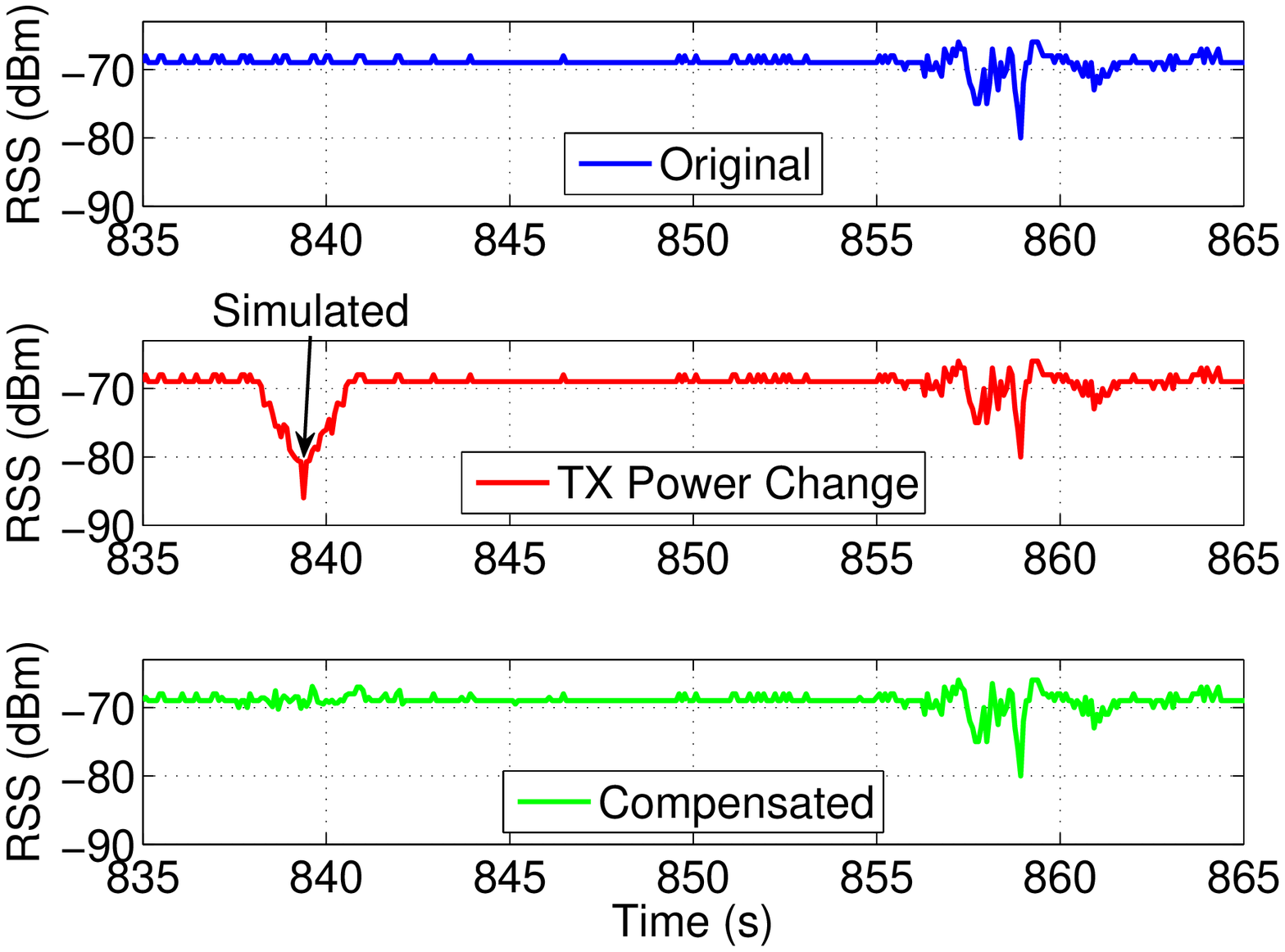,width=\columnwidth}}
        }
    \caption{Measured CSI and RSS (top) without and (middle) with TX power change; and (bottom) after compensation, which nullifies the effect of TX power change.}
  \label{fig:tx_comp}
  \end{center}
\end{figure*}

\subsection{Compensation for Transmit Power Change}
\label{sec:tx_power_change}
In this section, we show how transmit power changes (random or strategic) affect line crossing detection
accuracy and how our compensation method nullifies the effect of such power changes.

Figure \ref{fig:tx_comp}(a) shows the effect of random transmit power changes on line crossing detection 
for a WiFi link between a single transmitter-receiver antenna pair that is crossed three times by a
 moving person.  The top figure corresponds to the case when there is no transmit power 
change. This figure clearly shows distinct short time periods of high variance in the CSI corresponding to the times when the person crosses the link. However, transmit power change masks 
these distinct short term variance regions and renders line crossing detection ineffective as can
 be seen in the figure in the middle. The bottom figure plots the CSI for the
 same link after compensating for the transmit power changes as described in Section 
\ref{sec:txchange}. Clearly, our compensation method almost nullifies the masking effect
of transmit power changes and the attacker can detect three line crossings (high
short term variance region) from the compensated signal.

Similarly, Figure \ref{fig:tx_comp}(b) shows how strategic power changes can be used to simulate link 
line crossings,  and how our compensation method eliminates these artificial
variations. The top figure plots the RSS in dBm for a ZigBee link that is crossed during the time interval 856-860 s. The figure in
the middle shows one additional line crossing (high variance region) introduced in the link by strategic transmit power changes during time interval 838-841 s. However, as seen from the bottom figure, our compensation method gets rid of the false alarm introduced by strategic power change and we can detect the original line crossing from the compensated signal.

\begin{figure*}
  \begin{center}
    \mbox{
        \subfigure{\epsfig{figure=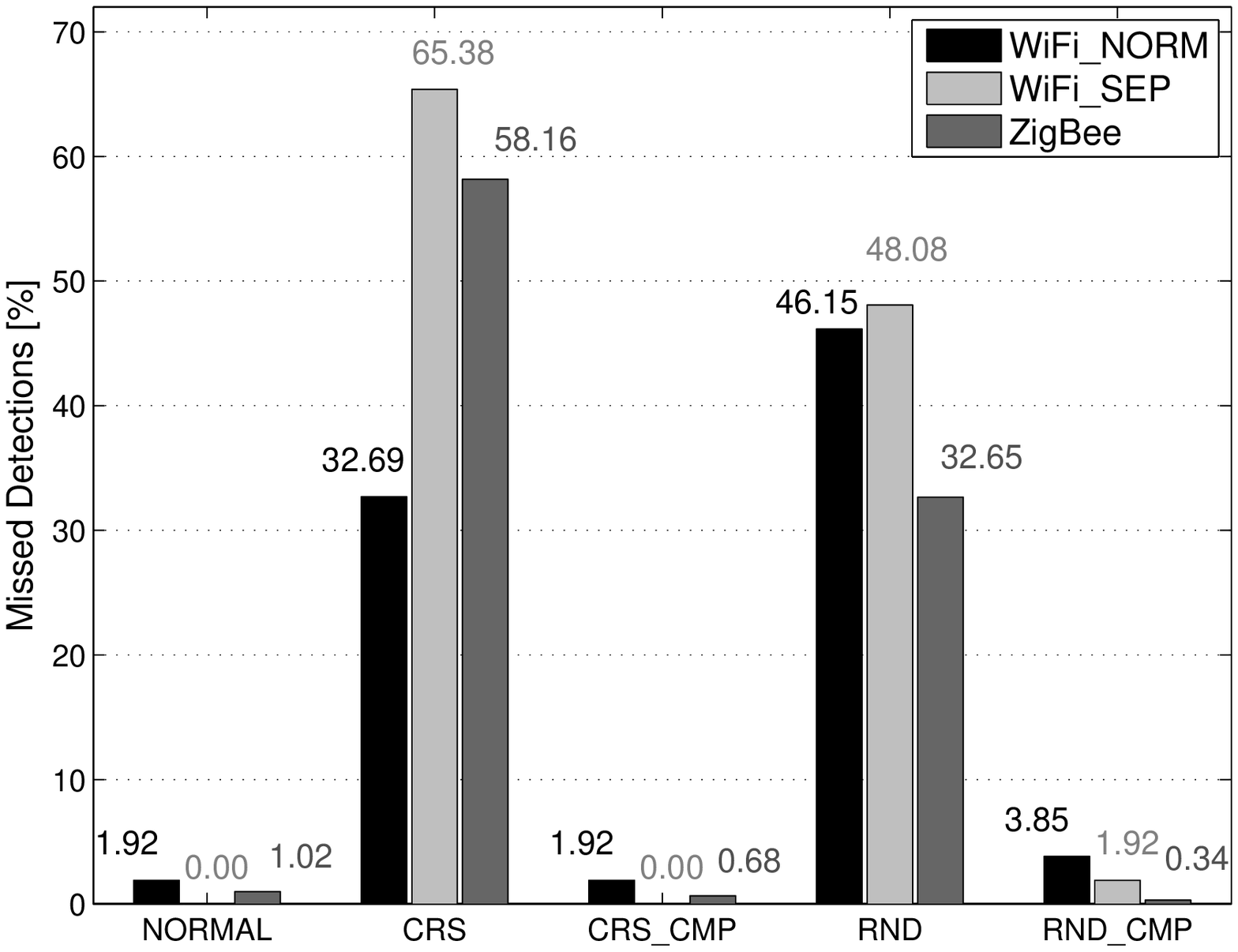,width=\columnwidth}} \quad
        \subfigure{\epsfig{figure=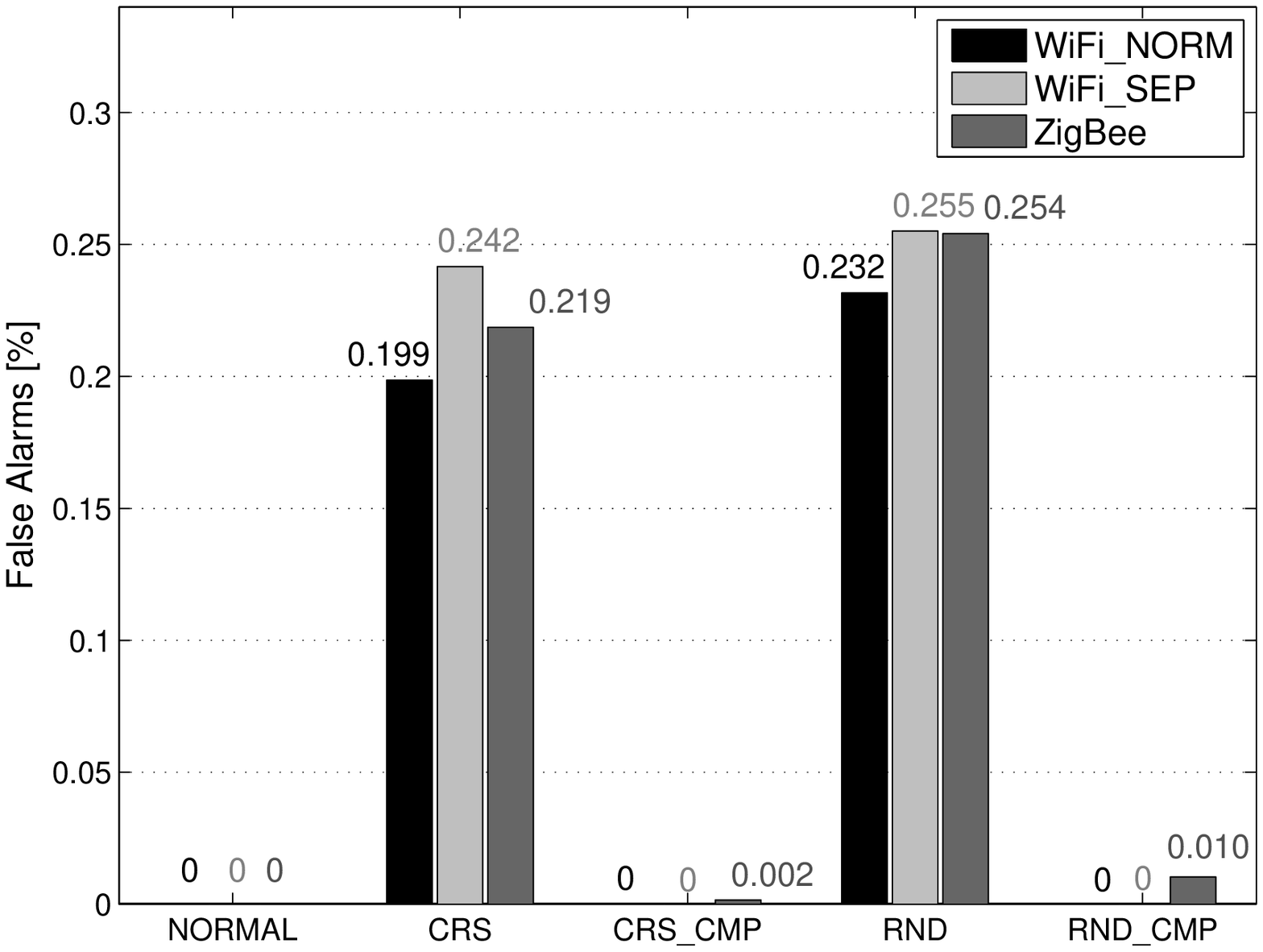,width=\columnwidth}}
        }
    \caption{Compensation accuracy in the University Hallway Experiment. Both strategic (CRS) and random (RND) transmit power variations increase missed detections and false alarms rate significantly. However, our
    compensation method eliminates most of these artificially induced missed detections and false alarms (see CRS\_CMP \& RND\_CMP).}
  \label{fig:fa_md_univ}
  \end{center}
\end{figure*}

\begin{figure*}
  \begin{center}
    \mbox{
        \subfigure{\epsfig{figure=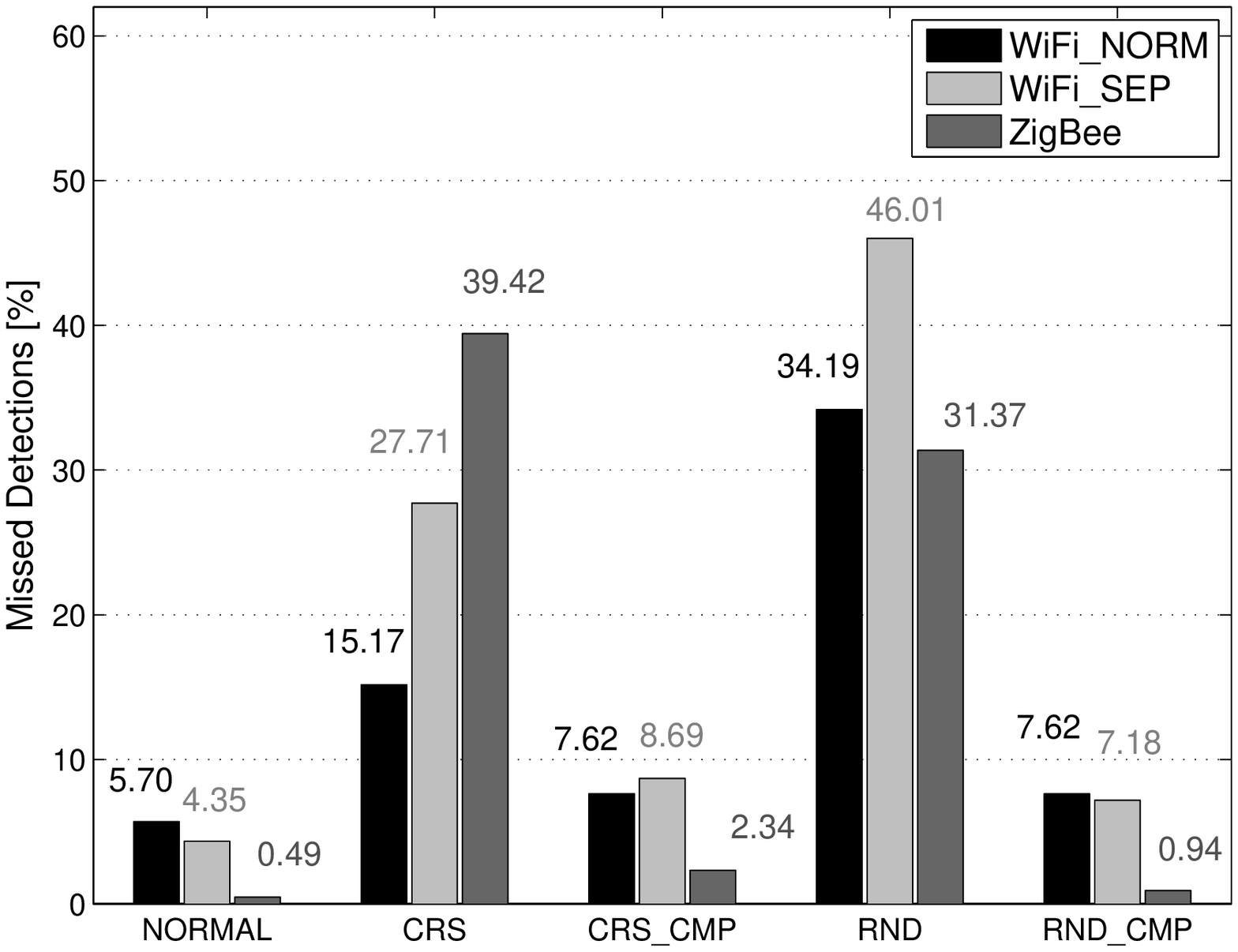,width=\columnwidth,height=2.5in}} \quad
        \subfigure{\epsfig{figure=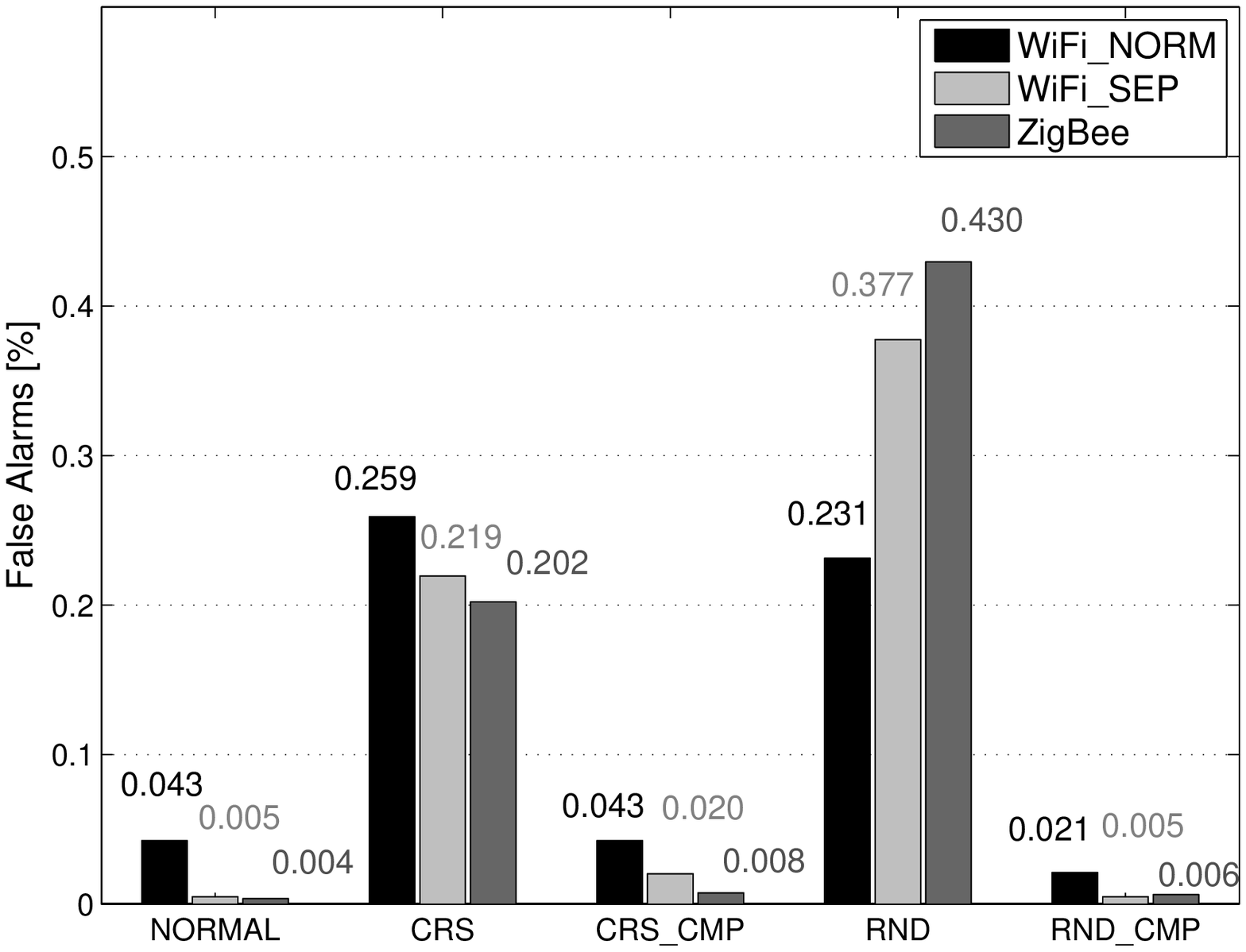,width=\columnwidth,height=2.5in}}
        }
    \caption{Compensation accuracy in the Residential House Experiment. Both strategic (CRS) and random (RND) transmit power variations increase missed detections and false alarms rate significantly. However, our compensation method eliminates most of these artificially induced missed detections and false alarms (see CRS\_CMP \& RND\_CMP).}
  \label{fig:fa_md_house}
  \end{center}
\end{figure*}

In the Figures \ref{fig:fa_md_univ} and \ref{fig:fa_md_house}, we show false alarms and missed detections induced by transmit power changes and the accuracy of our compensation method. In the figures, NORMAL
corresponds to the case when the transmitter transmits with fixed transmit power, CRS is when strategic power changes are introduced in the data using TX\_LINECROSS simulation, CRS\_CMP corresponds to the results when we apply our compensation method on TX\_CRS.
Similarly, RND shows results when the transmitter is changing its
transmit power randomly with each transmission, while RND\_CMP is the 
corresponding compensation results.
Note that the owner of the legitimate transmitter has full control over 
the transmitter node and can randomly select the periodicity with which to introduce transmit power changes in case of the TX\_LINECROSS experiment. We present results for one such simulated scenario where the owner randomly selects a time period between $3-10$ seconds to change transmit power according to a profile that mimics typical channel variation introduced by a person crossing the link line.  

We see that transmit power changes 
 (for both TX\_LINECROSS and TX\_RANDOM experiments) introduce significant false alarms and missed detections while using either WiFi (with or without spatially separated antennas) or ZigBee nodes. As an example, in the 
University Hallway experiment, a strategic 
transmit power change at the WiFi transmitter increases the missed detections rate from $1.92\%$ to $32.69\%$ and the false alarms rate from $0\%$ to $0.199\%$ when using a WiFi receiver with normal antenna separation. However, our compensation method gets rid of all the additional false alarms and missed detections.  
Similarly, in the Residential House experiment, for random power changes at the ZigBee transmitter,  
the missed detections rate increases to $31.37\%$ from $0.94\%$ and the false alarms rate increases to $0.429\%$
from $0.003\%$ but our compensation method brings down the missed detection and false alarm rates to only $0.94\%$ and $0.006\%$, respectively.

Using Equation \ref{eq:txest},
we can estimate the transmit power change amplitude accurately in $98\%$ cases
if we allow an error margin of $\pm2$ dB.

To summarize our findings, transmit power changes (strategic or random) increase the false alarm
and missed detection rates significantly. However, using our compensation method,
an attacker can accurately estimate the transmit power change amplitude and compensate 
for the same to get rid of most the adverse effect caused by such changes and still sense people 
location and motion with high accuracy.

\begin{figure*}
  \begin{center}
    \mbox{
        \subfigure{\epsfig{figure=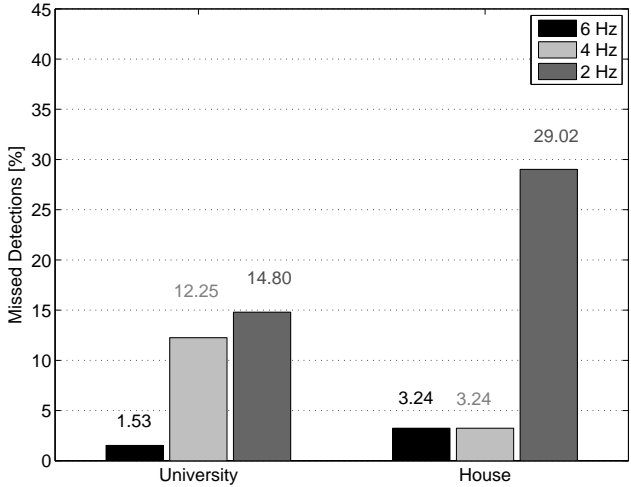,width=\columnwidth}} \quad
        \subfigure{\epsfig{figure=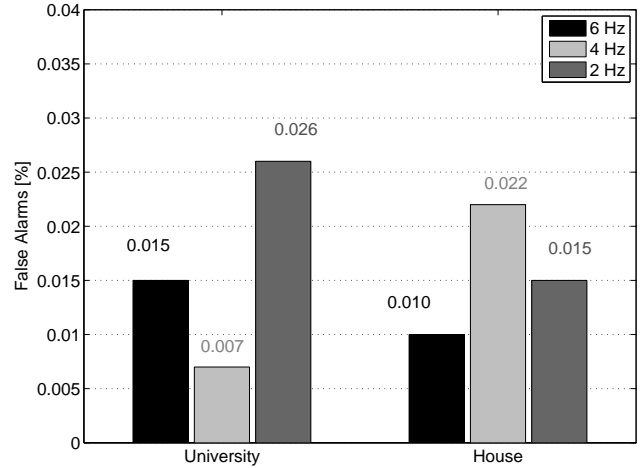,width=\columnwidth,height=2.5in}}
        }
    \caption{Detection accuracy with varying transmission rates (ZigBee).}
  \label{fig:diff_sample}
  \end{center}
\end{figure*}

\subsection{Detection with Varying Transmission Rate}
ZigBee applications in modern facilities use different transmission
rates for communication. 
In this section, we show how detection accuracy varies when the transmission rate for the ZigBee transmitter is lowered. We use
the data from TX\_NORMAL for both the University Hallway
and Residential House experiment to simulate the effect of 
lower transmission rate. Note that the original transmission rate
is approximately $12$ Hz. We simulate three additional transmission rates - $6$ Hz, $4$ Hz and $2$ Hz respectively from the original data. Figure \ref{fig:diff_sample} shows the results of our simulation. We find that the overall detection rates decrease with lower transmission rates. For the transmission rate of $6$ transmissions/second, accuracy of the detector 
is over $98\%$ for the University Hallway experiment and over $96\%$ for the Residential House experiment. These results are similar to what we observe for original transmission frequency of $12$ Hz. The accuracy is worst for transmission frequency of $2$ Hz with the detection rate being as low as $71\%$ for the Residential House experiment. For the transmission rate of $4$Hz, the detection rate degrades to $~87\%$ in the University Hallway experiment, although it remains above $96\%$ for the Residential House experiment.
We do not see any noticeable change in the false alarm rates with 
varying transmission rate.

We summarize our findings as follows:  Detection accuracy with ZigBee nodes decreases as transmission rate is lowered. For an ERW attack to succeed with high accuracy, the transmission frequency must be at least $6$ transmissions/second.

\section{Related Work}
\label{sec:related_work}

Preserving the privacy of the location of mobile devices in wireless networks has been object of intense research \cite{Bueller_2007,Danev_2012}. Location represents an important private information that can be used by malicious attackers for serious privacy violations and potentially dangerous attacks. The work in \cite{Rouf2010} presents an evaluation of the privacy and security of wireless tire pressure monitoring systems. It shows that eavesdropping these systems is possible through their static identifiers even at a distance of $40$ m.

Other works have demonstrated that communicating wireless devices leak the current and past location of people carrying these devices. In \cite{Rasmussen2008}, the authors show that distance bounding protocols \cite{dist_bound_protocols1993} can leak distance and location information to an attacker overhearing the communication between the prover and the verifier to such an extent as to allow the attacker to estimate his own position relative to the two devices. They also introduce a location private distance bounding protocol that protects against malicious provers, passive eavesdroppers, and attackers trying to actively initiate a distance bounding session. Fu et al. \cite{Fu2009} describe a system that can reveal the locations of WiFi-enabled mobile devices within the coverage area of a single high-gain antenna. By knowing the location and/or the maximum transmission range of the APs, an eavesdropper can set up a high-gain antenna to sniff the traffic between the \emph{victim} mobile device and the APs on all the available wireless channels and estimate the position of the mobile device. The work in \cite{Jiang2007} proposes three countermeasures to improve the location privacy in wireless networks, \emph{i.e.}, anonymize the identity of the device by frequently changing its pseudonym during communications (as in \cite{Gruteser2005}), un   link different pseudonyms of the same device with silent periods between different pseudonyms, reduce the transmission range of the devices through power control to minimize the number of APs that can collaborate to localize the devices' location (the precision to which a mobile device can be located depends on how many APs can hear from the device \cite{Bahl2000}).

The works focusing on location privacy typically assume that the \emph{victims} of the attack are carrying a wireless device (\emph{e.g.}, a mobile phone, RFID tag, low-power radio transceiver) that is actively communicating with the surrounding network infrastructure (\emph{e.g.}, WiFi APs, RFID readers, other radio transceivers). The work in \cite{Chetty_2012} presents a through-walls passive WiFi radar system. In it, a receiver is situated outside the target building and a Wi-Fi AP placed inside the building and having a narrow-beamwidth directional antenna is used as transmitter. The signal received by the passive radar detector is then used to create a range-Doppler surface and detect a moving target. Our work is complementary to \cite{Chetty_2012} because through wall radar systems have limited range due to direct signal interference. and further, as they are based on transmission, could be detected by source localization or counteracted by jamming. Other systems localize people by measuring the change in RSS of links traveling across an area where several WiFi APs or ZigBee radio transceivers are deployed. In the case of Wi-Fi based passive localization systems \cite{Nuzzer_2012}, a radio map of the environment is created by having a person standing at different locations while recording the RSS of all the links. This requires access to the target area for an initial calibration of the system. For radio tomographic systems \cite{VRTI}, accurate localization of people requires a high density deployment of radio transceivers on all sides of the target area.

In this work, we demonstrate that the presence, location and movements of people not carrying any wireless device can still be eavesdropped by measuring the RSS of the links between the devices composing the legitimate network and few receivers positioned outside the target area. This can be achieved without requiring a complex network infrastructure or previous access to the target area for an initial calibration. In \cite{manas_dspan}, the authors propose a method to detect an attack to a radio tomographic system in which some of the deployed radio transceivers are maliciously reprogrammed to change their transmit power. Our work is different in that we propose a method capable of correctly estimating the amplitude of the transmit power changes implemented by the legitimate devices as a countermeasure to the EPL attack. This enables reconstructing the true dynamics of the RSS signals and estimate people's locations. Moreover, in our work we do not make any assumption on the number of transmitters changing their transmit power and on the periodicity and amplitude of such changes.

\section{Conclusion and Future Work}
\label{sec:conclusion_and_future}
We investigated the ability of an attacker to surreptitiously use an otherwise secure wireless network to detect moving people through walls. We designed and implemented an attack methodology for through wall people localization that relies on reliably detecting when people cross the link lines by using physical layer measurements between the legitimate transmitters and the attack receivers. We also developed a method to determine the direction of movement of a person from the sequence of link lines crossed during a short time interval. Additionally, we described how an attacker may estimate any artificial changes in transmit power (used as a countermeasure), compensate for these power changes using measurements from sufficient number of links, and still detect line crossings. We implemented our methodology on WiFi and ZigBee nodes and experimentally evaluated the ERW attack by monitoring people movements through walls in two real-world settings.
 We found that our methods achieve close to $100\%$ accuracy in detecting line crossings and the direction of movement, when we use two WiFi 802.11n nodes with normal antenna separation, or two groups of ZigBee nodes as attack receivers. We also found that our methods achieve $90-100\%$ accuracy when we use a single 802.11n attack receiver.

Future work must develop more sophisticated protocols to prevent person location information leakage.  Device hardware enhancements may be necessary for this purpose.  

\balance
\bibliographystyle{abbrv}
\bibliography{radiowindow}

\end{document}